\documentstyle[epsf,amstex,amssymb,12pt]{article}
 1
 1
\newtheorem{pro}{Proposition} \newcommand{\dbar}{\bar{\partial}}
\newcommand{\bs}{{\boldsymbol{s}}} \newcommand{\bt}{{\boldsymbol{t}}}
\newcommand{\gr}{\operatorname{Gr}}

\begin{document}
 \title{\sc Singular sector of the KP hierarchy,
$\bar{\partial}$-operators of non-zero index and associated integrable
systems}
\author{Boris G. Konopelchenko \\ {\em Dipartimento di Fisica,
Universit\'a di Lecce} \\ {\em 73100 Lecce, Italy}
\\ Luis Mart\'\i nez
Alonso \\ {\em Departamento de F\'\i sica Te\'orica, Universidad
Complutense}\\ {\em E28040, Madrid, Spain}
\\ Elena Medina \\ {\em
Departamento de Matem\'aticas, Universidad de C\'adiz} \\{\em E11510
C\'adiz, Spain}} \date{} \maketitle

\begin{abstract} Integrable
hierarchies associated with the singular sector of the KP hierarchy, or
equivalently, with $\dbar$-operators of non-zero index are studied.  They
arise as the restriction of the standard KP hierarchy to submanifols of
finite codimension in the space of independent variables. For higher
$\dbar$-index these hierarchies represent themselves families of
multidimensional equations with multidimensional constraints. The
$\dbar$-dressing method is used to construct these hierarchies. Hidden
KdV, Boussinesq and hidden Gelfand-Dikii hierarchies are considered too.
\end{abstract} \newpage

\section{Introduction}

It is well established now that the Kadomtsev-Petviashvili (KP) hierarchy
is the key ingredient in a number of remarkable nonlinear problems, both
in physics and mathematics (see e. g. \cite{zm}-\cite{k1}). In physics,
its
applications range from the shallow water waves (see \cite{zm}-\cite{k1})
to the modern
string theory (see e.g. \cite{w1}-\cite{vm}). Resolution of the famous
Schottky problem
is one of the most impressive manifestation of the KP hierarchy in pure
mathematics
\cite{sh}.
Several methods have been developed to describe and analyze the KP
hierarchy
and other
integrable hierarchies, for instance, the inverse scattering transform
method \cite{zm}-\cite{k1}, Grassmannian approach \cite{sa}-\cite{sg}
or $\dbar$- dressing
method \cite{zm2}-\cite{z1},\cite{k1},\cite{bz}.
These methods have been rised to study generic properties of the KP
hierarchy and other integrable equations. In particular, the construction
of everywhere regular solutions of integrable systems (solitons, lumps,
dromions, ect),
which may have applications in physics, was of a main interest.

Much less attention was paid to singular solutions of integrable
equations. Pole type solutions of the Korteweg-de Vries (KdV) equation
have been known for a long time. However, the interest for
 this class of solutions
has increased only when it was shown that the motion of poles for
the KdV equation is governed by the Calogero-Moser model
\cite{amm},\cite{ch}. Similar results for rational singular
solutions of the KP equation have been obtained in \cite{kr}. The
general study of generic singularity manifolds began with the
formulation of the Painlev\'e  analysis method for partial
differential equations in \cite{wtc},\cite{w2}. Structure of
generic singularities of integrable equations has occured to be
connected with all their remarkable properties (Lax pairs,
Backlund-transformations, ect) (see e.g.\cite{rgb},\cite{ac}.)
Characteristic singular manifolds (i.e. singular manifolds with
additional constraints ) have been discussed in \cite{we},\cite{es}
and \cite{pi}.

A new method to analyze singular sectors of integrable equations
has been proposed in \cite{am1}. It uses the Birkhoff decomposition
of the Grassmannian, its relation with zero sets of the
$\tau$-function and its derivatives, and properties of Backl\"{u}nd
transformations. This methods provides a regular way to construct
desingularized wave functions near the blow-up locus (Birkhoff
strata). Note that the connection between the Painlev\'e analysis
and cell decomposition for the Toda lattice has been discussed in
\cite{am2} (see also (\cite{nt}). Note also that the characteristic
singular manifolds considered in \cite{we} and \cite{pi} correspond
to the second Birkhoff stratum ($\tau=0,\; \tau_x=0$).

\vspace{2mm}

A problem closely related with the study of these singular sectors is the
following: a standard KP (and KdV) hierarchy flows in the so-called big
cell of the
Grassmannian (dense open subset of the Grassmannian). The Birkhoff strata
are
subsets with finite codimension. Are there any integrable systems
associated with the Birkhoff strata? Positive answer to this question have
given recently in \cite{mmm}. It occurs that in the KdV case the
corresponding integrable
hierarchies are connected to the Schr\"{o}dinger equation with
energy-dependent potential.

In the present paper we study integrable hierarchies associated with the
singular sector
of the KP hierarchy. This sector consists of different Birkhoff strata
or equivalently of different Schubert cells. The Schubert cells have
finite
dimension  and are connected with
 the family of the Calogero-Moser type models
which describe motion of poles. Here we will concentrate on
integrable systems associated with Birkhoff strata. We show that
they can be constructed by restricting the standard KP hierarchy to
submanifolds of finite codimension in the space of independent
variables. To build these hierarchies  we will mainly use the
$\dbar$-dressing method. Integrable systems associated with
Birkhoff strata are rather complicated as well as the corresponding
linear problems. They are of high order, though there are
effectively 2+1 dimensional hierarchies. For higher Birkhoff strata
these integrable equations clearly demonstrate  a sort of quasi
multi-dimensionality. We also discuss hidden Gelfand-Dikii
hierarchies. Besides of illustrating by simplier formulae some of
the results of this work, we can in this case provide useful
methods to construct solutions.

An important property of the hidden KP hierarchies is that they are
associated with
$\dbar$-operators of non-zero index. This result is due to the
interpretation of the Grassmannian
as the space of boundary conditions for the $\dbar$-operator acting on the
Hilbert space
of square integrable functions. We prove that the codimension of Birkhoff
strata
coincides with the index of corresponding $\dbar$-operator up to sign.

\vspace{2mm}

We finish this introduction by describing the plan of the work. In Section
2
we remind some basic facts about the KP hierarchy, we briefly present the
$\dbar$-dressing method in subsection 2.1, the grassmannian and its
stratification
is reviewed in subsection 2.2 and the relation between singular sectors
of the KP
hierarchy and $\dbar$ operators of nonzero index is considered in
subsection 2.3.
Next, we devote Section 3 to the construction of the hidden KP
hierarchies,
first the case Index$\,\dbar=-1$ is carefully analyzed in subsection 3.1.
The case
Index$\,\dbar=-2$ is studied in subsection 3.2 and
the cases of  Index$\,\dbar=-3$ and higher indices are discussed in
subsection 3.3.
Finally, hidden Gelfand-Dikii hierarchies are analyzed in Section 4, we
prove
that under certain conditions the only hidden Gelfand-Dikii hierarchies
are
the KdV hidden hierarchy and the Boussinesq hidden hierarchy. The first
one
is studied in subsection 4.1, the second one in subsection 4.2. A
method of
constructing solutions is developed in subsection 4.3.

\section{The standard KP hierarchy and some general methods}

We start by remaining some basic facts about the KP hierarchy
and some of the methods developed to its study .
The KP hierarchy can be described in various ways
(see e.g. \cite{zm}-\cite{vm}).
The most compact form
of it is given by the Lax equation
\begin{equation}\label{2.1}
\frac{\partial L}{\partial t_n}=[(L^n)_+,L],\quad n=1,2,3,\dots,
\end{equation}
where
\begin{equation}\label{2.2}
L=\partial+u_1\partial^{-1}+u_2\partial^{-2}+u_3\partial^{-3}+\cdots
\end{equation}
is the formal pseudo-differential operator,
$\partial\equiv\frac{\partial}{\partial t_1}$
and $(L^n)_+$ denotes the differential part of $L^n$. Equation (\ref{2.1})
is an infinite set of equations for scalar functions $u_1$, $u_2$,
$u_3$,\dots.
These equations allow to express $u_2$, $u_3$,\dots via $u_1$ and its
derivatives with respects to
$t_1$ and $t_2$. As a result one gets the usual form of the KP hierarchy
given by equations
\begin{equation}\label{2.3}
\frac{\partial u}{\partial t_n}=f_n\left(u,\frac{\partial u}{\partial
t_1},
\frac{\partial u}{\partial t_2},\dots,\frac{\partial^{-1} u}{\partial
t_1},\dots\right),
\quad n=3,4,5,\dots,
\end{equation}
where $u=u_1$ and $f_n$ are certain functions on $u$,
$\frac{\partial u}{\partial t_1}$,
$\frac{\partial u}{\partial t_2}$,...,$\frac{\partial^{-1} u}{\partial
t_1}$,...
The simplest of these equations is
\begin{equation}\label{2.4}
\frac{\partial u}{\partial t_3}=\frac{\partial^3 u}{\partial t_1^3}+
6u\frac{\partial u}{\partial t_1}+
3\left(\frac{\partial  }{\partial t_1}\right)^{-1}
\frac{\partial^2 u}{\partial t_2^2}.
\end{equation}
Equations (\ref{2.1}) arise as the compatibility condition of the linear
equations
\begin{equation}\label{2.5}
L\psi=\lambda\psi\end{equation}
and
\begin{equation}\label{2.6}
\frac{\partial \psi}{\partial t_n}=(L^n)_+\psi,\quad n=1,2,3,\dots,
\end{equation}
where $\psi=\psi(\bt,\lambda)$ is the wave function of the KP hierarchy,
$\lambda$
is a complex parameter (spectral parameter) and
$\bt=(t_1,t_2,\dots,t_n,\dots)\in{\Bbb C}^{\infty}$. This wave function
 has the form
\begin{equation}\label{2.7}
\psi=e^{\sum_{n=1}^{\infty}\lambda^nt_n}\chi(\bt,\lambda)
\end{equation}
where
\begin{equation}\label{2.8}
\chi(\bt,\lambda)=1+\frac{\chi_1(\bt)}{\lambda}+\frac{\chi_2(\bt)}{\lambda^2}+
\cdots\quad \mbox{for large }\lambda.
\end{equation}
The functions $\psi(\bt,\lambda)$ and the adjoint wave function
$\psi^*(\bt,\lambda)$
(solution of equations formally adjoint to equations
(\ref{2.5}),(\ref{2.6}))
obey the famous Hirota bilinear equation
\begin{equation}\label{2.9}
\int_{S_{\infty}}d\lambda\,\psi(\bt,\lambda)\psi^*(\bt',\lambda)=0\quad
(\mbox{all }\bt\mbox{ and }\bt')
\end{equation}
where $S_{\infty}$ is a small circle around $\lambda=\infty$.

Finally, the wave-function is connected with the $\tau$-function
via
\begin{equation}\label{2.10}
\chi(\bt,\lambda)=\frac{\tau(\bt-[\lambda^{-1}])}{\tau(\bt)}
\end{equation}
where $[a]:=(a,\frac{1}{2}a^2,\frac{1}{3}a^3,\dots)$.

\subsection{The $\dbar$-dressing method}

Now, we present a sketch of the $\dbar$-dressing method
(see e.g.\cite{zm2}-\cite{z1}).
It is based on the nonlocal $\dbar$ problem:
\begin{equation}\label{2.12}
\frac{\partial\chi(\bt,\lambda,\bar{\lambda})}{\partial \bar{\lambda}}=
\int\!\!\!\int_Gd\mu\wedge d\bar{\mu}\chi(\bt,\mu,\bar{\mu})g(\bt,\mu)
R_0(\mu,\bar{\mu},\lambda,\bar{\lambda})g^{-1}(\bt,\lambda)
\end{equation}
where $\chi$ is a scalar function,
$R_0(\mu,\bar{\mu},\lambda,\bar{\lambda})$
is an arbitrary function, bar means complex conjugation, $G$ is a domain
in ${\Bbb C}$
and $g(\bt,\mu)$ is a certain function of $\bt$ and  the spectral
parameter.
It is assumed that $\chi$ is properly normalized
$(\chi\stackrel{\lambda\rightarrow\lambda_0}{\longrightarrow}
\eta(\lambda_0))$ and equation (\ref{2.12}) is uniquely
solvable. In virtue of the generalized Cauchy formula, the $\dbar$ problem
(\ref{2.12}) is equivalent
to a linear integral equation. The form of this linear equations (and
corresponding nonlinear equations,
associated with (\ref{2.12})) is encoded in the dependence of the function
$g$
on $\bt$. To extract these equations, we introduce long derivatives
\begin{equation}\label{2.13}
\nabla_n=\frac{\partial}{\partial
t_n}+g^{-1}(\bt,\lambda)g_{t_n}(\bt,\lambda)
\end{equation}
where $g_{t_n}\equiv\frac{\partial g}{\partial t_n}$. Then, we consider
the Manakov
ring of differential operators of the form
\begin{equation}\label{2.14}
L=\sum_{n_1,n_2,\dots}u_{n_1n_2n_3\cdots}(\bt)\nabla_1^{n_1}\nabla_2^{n_2}\nabla_3^{n_3}\cdots
\end{equation}
where $u_{n_1n_2n_3\cdots}(\bt)$ are scalar functions. In this ring we
select those $L$ which obey the
conditions
\begin{equation}\label{2.15}
\big[L,\frac{\partial}{\partial \bar{\lambda}}\big]\chi=0
\end{equation}
and $L\chi\rightarrow0$ as $\lambda\rightarrow\infty$. Condition
(\ref{2.15})
means that $L\chi$ has no singularities in $G$. The unique solvavility of
(\ref{2.12}) implies that for such $L$ one has
\begin{equation}\label{2.16}
L_{i}\chi=0.
\end{equation}
The set of equations (\ref{2.16}) is known as the system of linear
problems. Note that taking into account that
$\frac{\partial}{\partial t_n}\psi=g\nabla_n\chi$, equations
(\ref{2.16}) can be equivalently written as:
\begin{equation}\label{2.17}
L_{i}\psi=0\end{equation} where in operators $L_i$ one has to
substitute $\nabla_n$ by $\frac{\partial}{\partial t_n}$. The
compatibility conditions of (\ref{2.16}) (or (\ref{2.17})) are
equivalent to nonlinear equations for $u_{n_1n_2n_3\cdots}(\bt)$,
which are solvable by the $\dbar$-dressing method. One has to
select a basis among an infinite set of linear equations
(\ref{2.16}) (or (\ref{2.17})). If one consider an infinite family
of times $t_n$ $(n=1,2,3,\dots)$ one has an infinite basis of
operators $L_i$ and, consequently an infinite hierarchy of
nonlinear  integrable equation associated with (\ref{2.12}).

To get the standard KP hierarchy one can choose the canonical
normalization of $\chi$
(i.e. $\chi\rightarrow
1+\frac{\chi_1}{\lambda}+\frac{\chi_2}{\lambda^2}+\cdots$
as $\lambda\rightarrow\infty$) and put
$$g=\exp\big(\sum_{n=1}^{\infty}\lambda^nt_n\big),$$
the long derivatives $\nabla_n$ are $\nabla_n=\frac{\partial}{\partial
t_n}+\lambda^n$
$(n=1,2,3,\dots)$ and the corresponding linear problems take the form
\begin{equation}\label{2.18}
L_n\chi=\left(\nabla_n-\sum_{k=0}^{n}u_k(\bt)\nabla_1^k\right)\chi=0,
\end{equation}
or equivalently
\begin{equation}\label{2.19}
\frac{\partial\psi}{\partial t_n}=\sum_{k=0}^{n}u_k(\bt)
\frac{\partial^k\psi}{\partial t_1^k},\quad n=1,2,3,\dots,
\end{equation}
where $u_k(\bt)$ are scalar functions. Equations (\ref{2.19}) are just
equations
(\ref{2.6}) and their compatibility conditions are equivalent to the KP
hierarchy in the usual form.
The $\dbar$-dressing method provides a wide class of exact explicit
solutions of the KP hierarchy which correspond to degenerate kernels
$R_0(\mu,\bar{\mu},\lambda,\bar{\lambda})$ of the $\dbar$-problem
(\ref{2.12})
(see \cite{zm2}-\cite{z1},\cite{k1}).
It is worth to realize that the $\dbar$-problem for $\psi$ and the adjoint
$\dbar$ problem
for $\psi^*(\bt,\lambda)$ imply the Hirota bilinear identity (\ref{2.9}).

Note that in the KP case, the domain $G$ is $D_0={\Bbb C}-D_{\infty}$
where
$D_{\infty}$ is a small disk around $\lambda=\infty$
($\partial D_{\infty}=S_{\infty}$).
In a similar manner, one can formulate the KP hierarchy if one chooses
$G$ such that $\partial G=S$ (being $S$ the unit circle).

\subsection{Grassmannian and stratification}

Next, we comment some basic facts about the Grassmannian approach in
relation
to the standard KP hierarchy.
Following \cite{vm},\cite{am1}, we consider the Grassmannian $\gr$ as the
set
of linear subspaces $W$ of formal Laurent series on the circle
$S_{\infty}$.
That means that $W$ possesses an algebraic basis
\begin{equation}\label{3.1}
W=\{w_0(\lambda),w_1(\lambda),w_2(\lambda),\dots\}
\end{equation}
with the basis elements
\begin{equation}\label{3.2}
w_n(\lambda)=\sum_{i=-\infty}^{s_n}a_i\lambda^i
\end{equation}
of finite order. Here $s_0<s_1<s_2<\cdots$ and $s_n=n$ for large $n$.
It can be proved that $\gr$ is a connected Banach manifold which exhibits
a
stratified structure \cite{sg},\cite{am1}. To describe this structure one
introduces
the set ${\cal S}_0$ of increasing sequences of integers
\begin{equation}\label{3.3}
S=\{s_0,s_1,s_2,\dots\}
\end{equation}
such that $s_n=n$ for large $n$. One can associate to each
$W\in\gr$ the set of integers
$$S_W=\{n\in{\Bbb Z}:\;\;\exists w\in W\mbox{ of order }n\}\in {\cal
S}_0.$$
On the other hand, given
$S\in{\cal S}_0$ one may define the subset of $\gr$
\begin{equation}\label{3.4}
\Sigma_S=\{W\in\gr:\;\; S_W=S\}
\end{equation}
which is called the Birkhoff stratum associated with $S$. The stratum
$\Sigma_{S}$ is a submanifold of $\gr$ of finite codimension
$l(S)=\sum_{n\geq 0}(n-s_n)$. In particular, if $S=\{0,1,2,3,\dots\}$ the
corresponding stratum has codimension zero and it is a dense open subset
of $\gr$ which is called the principal stratum or the big cell.
Lower Birkhoff strata correspond to $S=\{s_0,s_1,s_2,\dots\}$ different
from $\{0,1,2,3,\dots\}$.

The KP hierarchy wave function $\psi(\bt,\lambda)$ (\ref{2.7}),
(\ref{2.8})
leads naturally to a family $W(\bt)$ in $\gr$ \cite{am1}. In order to see
it
 we start by introducing $(\lambda\in S_{\infty})$:
\begin{equation}\label{3.6}
W=\mbox{span}\{\psi(\bt,\lambda),\mbox{ all } \bt\}.
\end{equation}
Using (\ref{2.6}) and Taylor expansions, one gets
\begin{equation}\label{3.7}
W=\mbox{span}\{\psi,\partial_1\psi,\partial_1^2\psi,\dots\},
\end{equation}
where $\partial\equiv\frac{\partial}{\partial t_1}$. Now, the flow defined
as:
$$W(\bt):=e^{-\sum_{k\geq 1}\lambda^nt_n}W,$$
can be characterized as
\begin{equation}\label{3.8}
W(\bt)=\mbox{span}\{\chi(\bt,\lambda),\nabla_1\chi(\bt,\lambda),\nabla_1^2\chi(\bt,\lambda),
\dots\} \end{equation}
where $\nabla_1=\frac{\partial}{\partial t_1}+\lambda$. Since
$\nabla_1^n\chi=\lambda^n+O(\lambda^{n-1})$, one has
\begin{equation}\label{3.9}
S_{W(\bt)}=\{0,1,2,\dots\}.
\end{equation}
So the flows $W(\bt)$ generated by the standard KP hierarchy belong to the
principal Birkhoff stratum \cite{am1},\cite{go}. Then, it seems natural to
wonder if
there exist integrable structures associated to other Birkhoff strata.
In this sense, only recently some   progress has been
 made. In \cite{mmm} it was shown that for the KdV hierarchy,
(reduction of the KP hierarchy)
evolutions associated with the Birkhoff strata are given by integrable
hierarchies arising from
the Schr\"{o}dinger equations with energy dependent potentials.
One of the main goals of the present paper is the study of integrable
structures
associated with the full KP hierarchy outside the principal stratum.

\subsection{$\dbar$-operators of nonzero index and singular sector of KP}

Finally, we propose a wider approach which reveals the connection of
stratification of
the Grassmannian with the analytic  properties of the
$\dbar$ operators.
This approach is based on the observation that the Grassmannian can be
viewed
as the space of boundary conditions for the $\dbar$-operator \cite{w1}.
Let us consider the Hilbert space $H$ of square integrable functions
 $w=w(\lambda,\bar{\lambda})$
on $\Omega:={\Bbb C}-D_{\infty}$ (where $D_{\infty}$  is a small disk
around the point
$\lambda=\infty$), with respect to the bilinear form:
\begin{equation}\label{4.1}
\langle u,v\rangle=\int\!\!\!\int_{\Omega}u(\lambda,\bar{\lambda})
v(\lambda,\bar{\lambda})\frac{d\lambda\wedge d\bar{\lambda}}{2\pi
i\lambda}.
\end{equation}
Then, given $W\in\gr$ (described above) there is an associated domain
${\cal D}_W$ on $H$ for $\dbar$, given by those functions $w$ for which
$\dbar w\in H$
and such that their boundary values on $S_{\infty}$ are in $W$. Thus, we
have an
elliptic boundary value problem. To formulate it correctly, that is to
have a skew-symmetric $\dbar$ operator:
\begin{equation}\label{4.2}
\langle v,\dbar u\rangle=-\langle \dbar v,u\rangle\quad \forall u\in {\cal
D}_W,
\quad \forall v\in {\cal D}_{\tilde{W}}
\end{equation}
one has to define $\tilde{W}$, the dual of an element $W$ in the
Grassmannian,
as the space of formal Laurent series $v(\lambda)$ of $\lambda\in
S_{\infty}$
$(S_{\infty}=\partial D_{\infty})$ which obey the condition
\begin{equation}\label{4.3}
\int_{S_{\infty}}\frac{d\lambda}{2\pi i\lambda}v(\lambda)u(\lambda)=0,
\quad \forall u\in W.
\end{equation}
Properties of $W$ and $\tilde{W}$ are convenient to evaluate the index of
the
$\dbar$-operator. Let $S_W$ and $S_{\tilde{W}}$ be subsets of integers
determined by the
orders of elements in $W$ and $\tilde{W}$. Then, we have
\begin{equation}\label{4.4}
S_{\tilde{W}}=\{-n|n\not\in S_W\}.
\end{equation}
Let $\dbar_W$ denote the operator $\dbar$ acting on the domain ${\cal
D}_W$.
The index of this operator is defined as
$$\mbox{Index}\,\dbar_W:=\mbox{dim}(\mbox{ker}\,\dbar_W)-
\mbox{dim}(\mbox{coker}\,\dbar_W).$$
It can be determined as
\begin{equation}\label{4.5}
\mbox{Index}\,\dbar_W=\mbox{card}(S_W-{\Bbb N})-\mbox{card}(S_{\tilde{W}}-
{\Bbb N})
\end{equation}
where ${\Bbb N}\equiv\{0,1,2,\dots\}$. Note that the index of the
$\dbar_W$
operator
is closely connected to the notion of virtual dimension of $W$ used in
\cite{sg}.
$$\mbox{v.d.}(W)=\mbox{card}(S_W-{\Bbb N})-\mbox{card}({\Bbb
N}-S_{\tilde{W}}).$$
Indeed,
\begin{equation}\label{ind}
0\in S_W\Rightarrow \mbox{Index}\,\dbar_W=\mbox{v.d.}(W).
\end{equation}
Let us consider now subspaces $W(\bt)$ generated by the standard KP
hierarchy.
Since $S_{W(\bt)}=\{0,1,2,\dots\}$ (see (\ref{3.9})) and
$S_{\tilde{W}(\bt)}=\{1,2,\dots\}$, one has that
$\mbox{Index}\,\dbar_{W(\bt)}=0$.
Thus, all the equations of the standard KP hierarchy are associated with a
sector
of $\gr$ with zero index of $\dbar$.

Now, what about the case of nonzero index? How to characterize these
sectors in
$\gr$? Are there integrable systems associated with them? Addresing these
questions is the main subject of our paper. The answer can be formulated
as follows:
Given the wave function $\psi(\bt,\lambda)$ of the standard KP hierarchy,
we
consider submanifolds ${\cal M}$ of finite codimension in the space ${\Bbb
C}^{\infty}$
which are defined by $m$ constraints
\begin{equation}\label{4.6}
f_i(\bt)=0,\quad i=1,2,\dots m
\end{equation}
($f_i$ are some analytic functions) imposed on the independent variables
$\bt=(t_1,t_2,\dots)$. The point is that under appropriate conditions,
the restriction  $\psi_{\mbox{res}}$ of the wave function $\psi$ on ${\cal
M}$
determines families $W_{\mbox{res}}(\bs)$ in $\gr$ which correspond to
$\dbar$
operators of nonzero index. We will show that these sectors of $\gr$ are
associated with integrable hierarchies.

The nonzero $\dbar$-index sector of $\gr$ is closely connected with its
singular
sector. Indeed, suppose that the restriction of $\psi$ on the manifold
(\ref{4.6})
defines the corresponding family $W_{\mbox{res}}(\bs)$ such that
\begin{equation}\label{4.7}
S_{W_{\mbox{res}}(\bs)}={\Bbb N}-\{r_1,r_2,\dots,r_l\}.
\end{equation}
Then it is clear that
\begin{equation}\label{4.8}
\lambda^{-l}W_{\mbox{res}}(\bs)\end{equation}
are elements of $\gr$ with zero virtual dimension and,
consequently, the same holds for
$\lambda^{-l}W_{\mbox{res}}$.
Therefore, there is a $\tau$-function, $\tau(\bt)$, associated with
$\lambda^{-l}W_{\mbox{res}}$ such that
$$\tilde{\chi}(\bt,\lambda)=\frac{\tau(\bt-[\lambda^{-1}])}{\tau(\bt)}.$$
On the other hand, the elements of minimal order in (\ref{4.8}) are
$\lambda^{-l}\chi_{\mbox{res}}(\bs,\lambda)$. This means that the function
$\tilde{\chi}$ is singular on the
submanifold ${\cal M}$ or, equivalently
\begin{equation}\label{4.9}
\tau|_{\cal M}\equiv0.\end{equation}
Then, the submanifolds ${\cal M}$ leading to the non-zero index sector of
$\gr$ are zero manifolds of the
$\tau$-function of the KP hierarchy. This property is rather obvious
if one observes that
(\ref{4.8}) determines a domain for the $\dbar$ operator with a
non-trivial kernel. Consequently,
the determinant of the $\dbar$ operator, which is proportional to
$\tau(\bt(\bs))$ \cite{sg}, vanishes. Therefore, the constrained wave
function
$\psi_{\mbox{res}}(\bs,\lambda)$
is a regularization of the wave function $\tilde{\psi}(\bt,\lambda)$ on
the blow-up submanifold ${\cal M}$.

The above coincidence between a singular and non-zero $\dbar$ index
sectors
is an
important
feature of the KP hierarchy. It demonstrates close relation between
singular and non-zero
$\dbar$ index aspects of the integrable hierarchies.

 Note that integrable hierarchies with constrained independent variables
have been discussed in different context also in \cite{[30]},\cite{[31]}.

\setcounter{equation}{0}

\section{Hidden  KP hierarchies}

Now, we proceed to the construction of nonlinear systems
associated with manifolds of finite codimensions given by constraints
(\ref{4.6}). First, we realize that for ``good'' functions $f_i$, the
theorem of implicit function implies that one can solve equations
(\ref{4.6})
with respect to any $m$ variables, i.e., one can express $m$ variables
$t_{n_1}$, $t_{n_2}$,\dots,$t_{n_m}$ as functions of the others
$\bs=(\dots,t_n,\dots)$ $(n\not\in\{n_1,\dots,n_m\})$ in the form
\begin{equation}\label{5.1}
t_{n_i}=b_i(\bs),\quad i=1,\dots,m.\end{equation}
Formulae (\ref{5.1}) gives us the parametrization of the manifold given by
(\ref{4.6}) by the independent variables $\bs$. Since any set of $m$ times
can
be choosen as $t_{n_i}$ in (\ref{5.1}), one has an infinite number of
different parametrizations of the same manifold (\ref{4.6}).

The KP wave function $\psi(\bt,\lambda)$ restricted to the manifold
(\ref{5.1}), i.e., the function $\psi_{\mbox{res}}(\bs,\lambda)$ is
regularizable [24].
Since in what follows, we have to consider
$\psi_{\mbox{res}}(\bs,\lambda)$
and $\chi_{\mbox{res}}(\bs,\lambda)$ instead of $\psi$ and $\chi$ we will
omit the label res in both cases.
In order to construct restricted KP hierachies we start with the
$\dbar$-problem
(\ref{2.12}) for the regularized restricted function $\chi(\bs,\lambda)$.
This function
has canonical normalization, the corresponding $\dbar$-problem is uniquely
solvable and
\begin{equation}\label{5.2}
g(\bs,\lambda)=e^{\sum_{n=1}^{\infty}\lambda^nt_n(\bs)}.
\end{equation}
Then, one can use all the machinery of the $\dbar$-dressing method.

\subsection{The case Index$(\dbar)=-1$}

We start our study of the hidden KP hierarchies by considering the
simplest case $m=1$. Suppose first that we solve  the constraint
$f(\bt)=0$ with respect to $t_1$, then we have
\begin{equation}\label{5.3}
t_1=b_1(s_2,s_3,\dots),\end{equation}
\begin{equation}\label{5.4}
g(\bs,\lambda)=e^{\lambda b_1(s)+\sum_{n=2}^{\infty}\lambda^ns_n}
\end{equation}
and the long derivatives $\nabla_n$ are
\begin{equation}\label{5.5}
\nabla_n=\frac{\partial}{\partial s_n}+\lambda \frac{\partial b_1(\bs)}
{\partial s_n}+\lambda^n,\quad n=2,3,\dots.
\end{equation}
Since in this case the operator of the first order in $\lambda$ is missing
in the basis (\ref{3.8})
one has
$$S_{W_{\mbox{res}}(\bs)}={\Bbb N}-\{1\},$$
consequently, due to (\ref{4.4}) and (\ref{4.5})
\begin{equation}\label{5.6}
\mbox{Index}(\dbar_{W_{\mbox{res}}(\bs)})=-1.\end{equation}
In order to get the linear problems (\ref{2.17}) associated to
(\ref{5.4}),
one has to construct the operators $L$ of the form (\ref{2.14})
which obey conditions (\ref{2.15}) and $L\chi\rightarrow0$ as
$\lambda\rightarrow\infty$.
This is a tedious but straightforward calculation after which one gets the
infinite
set of linear problems
\begin{equation}\label{esp}
\partial_x^3 \psi-\partial_y^2 \psi=u_0
\psi+
u_1\partial_x \psi+u_2\partial_y \psi+
u_3\partial_x^2 \psi+u_4\partial_x\partial_y \psi,\end{equation}
and
\begin{equation}\label{5.8}\begin{array}{l}
\partial_{s_4} \psi-\partial_x^2 \psi=v_0
\psi
+v_1\partial_x \psi+v_2\partial_y \psi,\\
\partial_{s_5}\psi=p_0\psi+p_1\partial_x \psi+p_2\partial_y \psi+
p_3\partial_x^2 \psi+\partial_x \partial_y\psi,\\
\dots\\
\partial_{s_n} \psi=\partial_x^{n_2}\partial_y^{n_3}\psi+\cdots.
\end{array}
\end{equation}
where $n=4,5,6,\dots$; $n=2n_2+3n_3$ with $n_2=2,3,\dots$ and $n_3=0,1$,
and we have made $x:=s_2$, $y:=s_3$. By denoting $b:=b_1$, we have that
the coefficients in (\ref{esp}), (\ref{5.8}) are given by
\begin{equation}\label{5.9}\begin{array}{l}
u_0= b_{xxx}\chi_1+3b_{xx}\chi_{1x}+3\chi_{2xx}+3b_x\chi_{1xx}+
3b_x^2\chi_{2x}+3b_x\chi_{3x}\\\\
\;\;+3\chi_{4x}-b_{yy}\chi_1-2b_y\chi_{1y}-3b_xb_{xy}\chi_1+b_xb_y\chi_{1x}
 \\\\
\;\;-3b_x^2\chi_{1y}-3b_x\chi_{2y}+4b_y\chi_{2x}-b_yb_{xx}\chi_1-3b_y\chi_1\chi_{1x}
\\\\
\;\;-3\chi_3\chi_{1x}-3b_x^2b_{xx}\chi_1-3b_x\chi_1\chi_{2x}
+2b_x\chi_1\chi_{1y}\\\\
\;\;-3b_x\chi_2\chi_{1x}-3\chi_2\chi_{2x}+3\chi_1\chi_2\chi_{1x}
+2\chi_2\chi_{1y}-2\chi_{3y},\\\\\\
u_1=3b_xb_{xx}+3b_x\chi_{1x}+3\chi_{2x}-3\chi_{1x}\chi_1-2\chi_{1y}-b_y^2-b_x^2b_y,\\\\
u_2=3b_{xx}+b_x^3+3\chi_{1x}+b_xb_y,\;\;u_3=-2b_y,\;u_4=3b_x,\\\\
v_0=2\chi_1\chi_{1x}-2\chi_{2x}-2b_x\chi_{1x},\;\;v_1=-b_x^2,\;\;v_2=-2b_x.\\\\
p_3=-b_x,\;\; p_2=2b_x^2-b_y,\\\\
p_1=b_x^3-b_xb_y-\chi_{1x},\\\\
p_0=\chi_1\chi_{2x}+\chi_1\chi_{1y}-2b_x\chi_1\chi_{1x}-\chi_1^2\chi_{1x}
+\chi_{1x}\chi_2-\chi_{3x}
\\\\
\;\;-\chi_{2y}-b_y\chi_{1x}-b_x\chi_{1y}+2b_x\chi_{2x}+2b_x^2\chi_{1x}.
\end{array}
\end{equation}
The absence of a operator $\nabla$ of the pure first order $\lambda$
imposes the constraints:
\begin{equation}\label{5.10}\begin{array}{l}
b_{xxx}+3\chi_{1xx}+3\chi_{3x}-b_{yy}-2\chi_{2y}-3b_xb_{xy}-b_x\chi_{1y}+
b_y\chi_{1x}\\\\
\;\;-b_{xx}b_y-3\chi_{1x}\chi_2-3b_x^2b_{xx}-3\chi_{2x}\chi_1+
3\chi_{1x}\chi_1^2+2\chi_{1y}\chi_1=0,\\\\
b_{s_4}-b_{xx}-2\chi_{1x}+2b_xb_y+b_x^3=0,\\\\
b_{s_5}-b_{xy}+b_xb_{xx}+b_y^2-b_x^2b_y-b_x^4-\chi_{2x}-\chi_{1y}\\\\
\;\;+3b_x\chi_{1x}+\chi_1\chi_{1x}=0
\end{array}
\end{equation}
and so on. Higher linear problems and the expresions of
the coefficients are complicated and we omit them. The linear problems
(\ref{esp})-(\ref{5.8}) represent themselves an infinite hierarchy of
the linear problems for the restricted (hidden) KP hierarchy with
Index$\,\dbar=-1$.
All these problems are compatible by construction and the compatibility
conditions for them
give rise to an infinite hierarchy of nonlinear evolution equations,
a restricted (hidden) KP hierarchy. The simplest system of this hidden KP
hierarchy is
the one associated with (\ref{esp}) and the first equation in (\ref{5.8})
and
 has the form ($t:=s_4$):
\begin{eqnarray}\label{5.11}
u_{4t}&=&-u_{4xx}-\frac{8}{9}u_4u_{4y}-\frac{5}{9}u_4^2u_{4x}+
\frac{4}{3}u_3u_{4x}+2u_{2x},\nonumber\\
 & & \nonumber   \\
u_{3t}&=&
-\frac{2}{3}u_4u_{4xx}-\frac{2}{3}u_{4x}^2+3v_{0x}+u_{3xx}+2u_{1x}-
\frac{1}{9}u_4^2u_{3x} \nonumber\\
 & & \nonumber\\
 & & -\frac{4}{3}u_3u_{4y}-\frac{2}{3}u_4u_{3y}+\frac{2}{9}u_4^2u_{4y}
  -\frac{2}{9}u_3u_4u_{4y},\nonumber\\
 & & \nonumber\\
u_{2t}&=&-\frac{2}{3}u_{4xxx}+u_{2xx}-\frac{1}{9}u_4^2u_{2x}-\frac{2}{3}u_2u_{4y}
+\frac{2}{3}u_{4yy}\nonumber\\
 & &\nonumber\\
 & &-\frac{2}{3}u_{4}u_{2y}-2v_{0y}+\frac{2}{3}u_4u_{4xy}-
 \frac{2}{3}u_2u_4u_{4x}\nonumber\\
 & &\nonumber  \\
 & &-u_4v_{0x}+\frac{2}{3}u_3u_{4xx}+\frac{2}{3}u_1u_{4x}, \\
 & & \nonumber\\
u_{1t}&=&-\frac{2}{9}u_4u_{4xxx}-\frac{8}{9}u_{4x}u_{4xx}+3v_{0xx}+u_{1xx}
+2u_{0x}\nonumber\\
 & & \nonumber\\
 &
&-\frac{1}{9}u_4^2u_{1x}-\frac{4}{3}u_1u_{4y}+\frac{2}{9}u_4u_{4yy}+\frac{2}{9}u_{4y}^2
-\frac{2}{3}u_4u_{1y}\nonumber\\
 & &  \nonumber\\
 & &-u_4v_{0y}+\frac{2}{9}u_4^2u_{4xy}+\frac{2}{9}u_4u_{4x}u_{4y}-
\frac{4}{9}u_1u_4u_{4x}\nonumber\\
 & & \nonumber\\
 &
&-\frac{2}{9}u_3u_4u_{4xx}-\frac{2}{9}u_3u_{4x}^2-2u_3v_{0x}+\frac{2}{9}u_2u_4u_{4y},\nonumber\\
 & &  \nonumber\\
u_{0t}&=&v_{0xxx}+u_{0xx}-\frac{1}{9}u_4^2u_{0x}-\frac{4}{3}u_0u_{4y}
-v_{0yy}-\frac{2}{3}u_4u_{0y}\nonumber\\
 & &-u_4v_{0xy}-\frac{2}{3}u_0u_4u_{4x}-u_3v_{0xx}-u_2v_{0y}-u_1v_{0x}
 \nonumber\end{eqnarray}
where
\begin{eqnarray}
v_0&=&\frac{1}{27}u_4^2u_3-\frac{1}{6}u_3^2+\frac{2}{9}u_4u_{4x}-\frac{2}{3}u_1
+\frac{4}{9}u_{4y}\nonumber\\
 & & \nonumber\\
 & &-\frac{4}{9}\partial_y\partial_x^{-1}\left(u_2-\frac{1}{27}u_4^3
 +\frac{1}{6}u_3u_4\right).\nonumber\end{eqnarray}
Note that since due to (\ref{5.9}) $u_4=3b_x$ and $u_3=2b_y$, one can
rewrite the
system (\ref{5.11}) as a system of four equations for the variables $b$,
$u_2$, $u_1$ and $u_0$. Equations (\ref{5.11}) and higher equations are
solvable
by the $\dbar$-dressing method, though all solutions would be expressed in
implicit
form. Then, we have an infinite hierarchy of integrable 2+1 dimensional
equations. This hierarchy is associated with  a restricted element in the
Grassmannian
which satisfies
\begin{equation}\label{5.12}
W_{\mbox{res}}(\bs)=\mbox{span}\{\nabla_x^n\nabla_y^m\chi(\bs,\lambda),
n\geq0,m=0,1\}.
\end{equation}
 Note
that the basis of the space
$W_{\mbox{res}}(\bs)$ is formed by  two-dimensional jets of special
form in contrast to the one-dimensional jets (\ref{3.8}) for the standard
KP hierarchy with
Index$\,\dbar=0$.

Note also that if one tries to construct the linear problem starting with
$g$ of the
form (\ref{5.4}) with $b_1=const.$, the procedure collapses. In this case
(\ref{5.10}) gives too strong constraints on the function $\chi$
($\chi_{1x}=0$,
$\chi_{2x}+\chi_{1y}=0$,...).

As it has been mentined before, the formula (\ref{5.3}) gives only one
possible parametrization
of the manifold defined by the equation $f(\bt)=0$. Let us take the one
 given by
\begin{equation}\label{6.1}
t_2=b_2(s_1,s_3,s_4,\dots), \end{equation}
so
\begin{equation}\label{6.2}
g(\bs,\lambda)=e^{\lambda s_1+\lambda^2b_2(\bs)+\sum_{i=3}^{\infty}
\lambda^is_i} \end{equation}
and long derivatives are
\begin{equation}\label{6.3}
\nabla_n=\frac{\partial}{\partial s_n}+\lambda^2\frac{\partial b_2}
{\partial s_n}+\lambda^n,\quad n=1,3,4,\dots.\end{equation}
Similar to the previous case, one gets an infinite family of linear
problems with the corresponding family
of constraints. The first linear problem is of the form ($x:=s_1$,
$y:=s_3$, $t:=s_4$)
$$\partial_x^3\psi=u_5\partial_y^2\psi+u_4\partial_x\partial_y\psi+
u_3\partial_x^2\psi+u_2\partial_y\psi+u_0\psi,$$
$$\partial_t\psi=v_3\partial_x^2\psi+v_2\partial_y\psi+v_1\partial_x\psi+
v_0\psi.$$
The compatibility condition for this system leads to the first integrable
system
of this hidden KP hierarchy. The expressions of the potentials, the
constraints
and the system are a bit more complicated than in the case studied above
and
we omit them here. Note that in this case we have
\begin{equation}\label{6.7}
W_{\mbox{res}}(\bs)=\mbox{span}\{\nabla_x^n\nabla_y^m\chi(\bs,\lambda),n\geq0,m=0,1\}
\end{equation}
and consequently
\begin{equation}\label{6.8}
\mbox{Index}(\dbar_{W_{\mbox{res}}(\bs)})=-1.\end{equation}
Let us consider now next choice
\begin{equation}\label{6.9}
t_3=b_3(s_1,s_2,s_4,\dots).\end{equation}
One has
\begin{equation}\label{6.10}
g=e^{\lambda
s_1+\lambda^2s_2+\lambda^3b_3(\bs)+\sum_{i=4}^{\infty}\lambda^is_i}
\end{equation}
and the long derivatives have the form
\begin{equation}\label{6.11}
\nabla_n=\frac{\partial}{\partial s_n}+\lambda^3\frac{\partial b_3}
{\partial s_n}+\lambda^n,\quad n=1,2,4,5,\dots.\end{equation}
Using these long derivatives we can find the corresponding hierarchy of
linear problems and nonlinear
integrable equations. On the first sight, one could think that as
$\nabla_x:=\nabla_1$ and $\nabla_y:=\nabla_2$ are third-order
operators, the operators of the first and the second order are missing
 and
consequently Index$\,\dbar=-2$. However, by taking the linear combination
of
long derivatives ($b:=b_3$ here)
$$b_y\nabla_x-b_x\nabla_y=b_y\partial_x+\lambda b_y-b_x\partial_y-
\lambda^2 b_x,$$
we see that only the first order in $\lambda$ is missing and then, we have
Index$\dbar=-1$.

Similar situation takes place in the general case
\begin{equation}\label{6.12}
t_n=b_n(s_1,\dots,s_{n-1},s_{n+1},\dots).\end{equation}
for $n\geq 3$. One has
\begin{equation}\label{6.13}
\nabla_m=\frac{\partial}{\partial s_m}+\lambda^n\frac{\partial b_n}
{\partial s_m}+\lambda^m,\quad m\neq n.\end{equation}
In a way similar to the previous case, one gets the operators of the
second, third,
..., $(n-1)-{th}$ order by taking linear combinations of the operators
$\nabla_1$,
$\nabla_2$,..., $\nabla_{n-1}$. Then, we have that in the general case
Index$\,\dbar=-1$.

We finally point out that all the hierarchies of linear problems  and
integrable systems considered in this subsection are closely connected.
In fact, since they are associated with different parametrizations
of the same manifold $f(\bt)=0$, they are related to each other by
change of independent and dependent variables.

\subsection{The case Index$(\dbar)=-2$}

Suppose now that we take $m=2$ in (\ref{4.6}), i.e. the manifold ${\cal
M}$
is defined as
$$f_1(\bt)=0,\quad f_2(\bt)=0.$$
Then, one possible parametrization is
$$\begin{array}{l}
t_1=b_1(s_3,s_4,\dots),\\
t_2=b_2(s_3,s_4,\dots),\end{array}$$
and the function $g$ and the long derivatives are given by
$$\begin{array}{l}
g(\bs,\lambda)=e^{\lambda b_1(\bs)+\lambda^2 b_2(\bs)+
 \sum_{i=3}^{\infty}\lambda^is_i},\\\\
\nabla_k=\displaystyle\frac{\partial}{\partial s_k}+\lambda \frac{\partial
b_1}
{\partial s_k}+\lambda^2\frac{\partial b_2}
{\partial s_k}+\lambda^k,\quad k=3,4,5,\dots.\end{array}$$
In order to construct linear problems we look for the operators of the
form
$$L=\sum u_{n_3\,n_4\,n_5\dots}\nabla_3^{n_3}\nabla_4^{n_4}\nabla_5^{n_5}
\dots$$
satisfying $L\chi=0$. It is easy to see that for $n\neq1,2$ we have an
order $n$ operator of the form
$\nabla_3^{n_3}\nabla_4^{n_4}\nabla_5^{n_5}$ with $n_3=0,1,2,\dots$ and
$n_4,n_5=0,1$. On the other hand, as each long derivative is
of a order
$\geq3$, there are not operators of order 1 and 2. Then,
\begin{equation}\label{i2.1}\begin{array}{l}
W_{\mbox{res}}(\bs)=
\mbox{span}\{\nabla_3^{n_3}\nabla_4^{n_4}\nabla_5^{n_5}\chi(\bs,\lambda),
n_3=0,1,2,\dots,n_4,n_5=0,1\}\\\\
S_{W_{\mbox{res}}(\bs)}={\Bbb N}-\{1,2\}\end{array}
\end{equation}
and consequently
$$\mbox{Index}(\dbar_{W_{\mbox{res}}(\bs)})=-2.$$
In order to get the linear problems involving the minimum number of
independent variables  we use, instead of (\ref{i2.1}), the more
convenient
system of generators of $W_{\mbox{res}}(\bs)$ given by
\begin{equation}\label{i2.2}
W_{\mbox{res}}(\bs)=
\mbox{span}\{\nabla_3^{n_3}\nabla_4^{n_4}\chi(\bs,\lambda),
n_3,n_4=0,1,2,\dots,\nabla_5\chi(\bs,\lambda)\}\end{equation}
The linear problems corresponding to the lowest orders are then
\begin{equation}\label{7.6}\begin{array}{l}
\partial_4^3\psi-\partial_3^4\psi+u_{11}\partial_3\partial_4^2\psi+
u_{10}\partial_3^2\partial_4\psi+u_9\partial_3^3\psi
+u_8\partial_4^2\psi+u_7\partial_3\partial_4\psi\\
   \\
\;\;+u_6\partial_3^2\psi+u_5\partial_5\psi
+u_4\partial_4\psi+u_3\partial_3\psi+u_0\psi=0,
\end{array}\end{equation}
\begin{equation}\label{7.7}
\partial_3\partial_5\psi-\partial_4^2\psi+v_7\partial_3\partial_4\psi
+v_6\partial_3^2\psi
+v_5\partial_5\psi+v_4\partial_4\psi+v_3\partial_3\psi+v_0\psi=0,
\end{equation}
\begin{equation}\label{7.8}\begin{array}{l}
\partial_5\partial_4\psi-\partial_3^3\psi+p_8\partial_3\partial_5\psi+
p_7\partial_3\partial_4\psi+p_6\partial_3^2\psi
+p_5\partial_5\psi+p_4\partial_4\psi\\
   \\
+p_3\partial_3\psi+p_0\psi=0,
\end{array}\end{equation}
where $u_i$, $i=0,3,4,\dots,11$, $v_k$ $k=0,3,\dots,7$ and $p_l$,
$l=0,3,\dots,8$
can be expressed in terms of $b_1$, $b_2$ and $\chi_n$, $(n\geq1)$. Due to
the
absence of operators of the first and second order in $\lambda$ we have
two
constraints on $b_1$, $b_2$ and $\chi_n$, $(n\geq1)$ associated to each
equation
(\ref{7.6})-(\ref{7.8}).

Note that only two linear equations among (\ref{7.6})-(\ref{7.8}) are
independent (form the basis of the Manakov ring). For instance, the
problems
(\ref{7.7}) and (\ref{7.8}) are equivalent modulo the problem (\ref{7.6}).
Indeed, acting by operator $\partial_4$ on (\ref{7.7}) and by the operator
$\partial_3$
on (\ref{7.8}), substracting the equations obtained and using (\ref{7.6})
one gets an identity. Then, choosing, for instance, (\ref{7.6}) and
(\ref{7.7})
we have a system of two three-dimensional linear equations
(variables $s_3$, $s_4$, $s_5$). However, it implies that $\psi$ satisfies
also a two-dimensional linear equation. Indeed, using (\ref{7.6}), one can
express
$\partial_5\psi$ via the derivatives of $\psi$ with respect to $s_3$ and
$s_4$ (since in general $u_5\not\equiv0$). By substituting this expression
into
(\ref{7.7}), one gets
\begin{equation}\label{7.9}\begin{array}{l}
\partial_3\partial_4^3\psi-\partial_3^5\psi+r_{14}\partial_3^2\partial_4^2\psi
+r_{13}\partial_3^3\partial_4\psi
+r_{12}\partial_4^3\psi+\tilde{r}_{12}\partial_3^4\psi\\
    \\
\;\;+r_{11}\partial_3\partial_4^2\psi+r_{10}\partial_3^2\partial_4\psi
+r_9\partial_3^3\psi+r_8\partial_4^2\psi+r_7\partial_3\partial_4\psi\\
   \\
\;\;+
r_6\partial_3^2\psi+r_4\partial_4\psi+r_3\partial_3\psi+r_0\psi=0,
\end{array}\end{equation}
where $r_i$ $(i=0,3,4,6,7,\dots,14)$ and $\tilde{r}_{12}$ can be expressed
in
terms of $u_i$ $(i=0,3,4,\dots,11)$ and $v_j$ $(j=0,3,4,\dots,7)$. Thus,
one
has
the two-dimensional equation (\ref{7.9}) with variables $s_3$ and $s_4$
and
the three-dimensional equation (\ref{7.6}). The compatibility condition
of the linear problem constituted by these two equations (or equivalently
(\ref{7.6}) and (\ref{7.7})) gives rise to a three-dimensional
nonlinear integrable system with independent variables $s_3$, $s_4$,
$s_5$.

In the same way, taking into account (\ref{i2.2}), the equations in
linear problems
which involve higher times $s_n$ $(n=6,7,\dots)$ can be written in the
form
\begin{equation}\label{7.10}
\frac{\partial\psi}{\partial s_k}=\partial_3^{n_3}\partial_4^{n_4}\psi+
\cdots+p_{k\,5}\partial_5\psi+\cdots+p_{k\,0}\psi,
\quad k=6,7,\dots\end{equation}
where $k=3n_3+4n_4$ and $p_{k\alpha}$ are certain functions and we have
two
constraints associated to each equation (\ref{7.10}).
In
order to eliminate $\partial_5\psi$ in (\ref{7.10}),we use (\ref{7.6})
so that
$$\frac{\partial\psi}{\partial s_k}=\partial_3^{n_3}
\partial_4^{n_4}\psi+
\sum_{m_3,m_4}\tilde{p}_{m_3\,m_4}(\bs)\partial_3^{m_3}
\partial_4^{m_4}\psi
\quad k=6,7,\dots.$$
By construction, these last equations are compatible  with equation
(\ref{7.9}),
thus, we have an infinite hierarchy of commuting 2+1-dimensional
integrable
systems for the coefficients.

Finally,  from (\ref{7.6}) we can express
$\nabla_5\chi$ as a linear
combination of elements of the form $\nabla_3^{n_3}\nabla_4^{n_4}\chi$, we
can
eliminate $\nabla_5\chi$ for the system of generators of
$W_{\mbox{res}}(\bs)$
(see (\ref{i2.1}), (\ref{i2.2})). In fact, by using
(\ref{7.6})-(\ref{7.8})
in (\ref{i2.1}) we get
$$
W_{\mbox{res}}(\bs)=
\mbox{span}\{\nabla_3^{n_3}\nabla_4^{n_4}\chi(\bs,\lambda),
n_3=0,1,2,\dots,n_4=0,1,2\}.$$

\subsection{Higher indices of the $\dbar$ operator}

We finish this section by discussing the basic properties of the case
Index$\,\dbar\leq-3$ (or equivalently $m\geq3$ in (\ref{4.6})). We will
show that we have a hierarchy of integrable systems associated to each
manifold defined by (\ref{4.6}) with $m\geq3$, but in this case the
hierarchy consists of 3+1 dimensional nonlinear systems, instead of
the 2+1 dimensional systems found in the cases Index$\,\dbar=-1$ and
Index$\,\dbar=-2$.
As all the basic properties are exhibited for $m=3$, we start by
considering
this particular case. Suppose then, that we take $m=3$ in (\ref{4.6}) and
solve
the constraints with respect to $t_1$, $t_2$ and $t_3$. We have
$$ t_1=b_1(\bs),\quad t_2=b_2(\bs),\quad t_3=b_3(\bs),$$
consequently
$$g(\bs,\lambda)=e^{\lambda b_1(\bs)+\lambda^2 b_2(\bs)+\lambda^3b_3(\bs)+
\sum_{i=4}^{\infty}\lambda^is_i},$$
and the long derivatives are
$$\nabla_k=\frac{\partial}{\partial s_k}+\lambda \frac{\partial b_1}
{\partial s_k}+\lambda^2\frac{\partial b_2}{\partial s_k}
+\lambda^3\frac{\partial b_3}{\partial s_k}+\lambda^k,\quad
k=4,5,\dots. $$
>From the expresions of the long derivatives, it is easy to see that, for
$n\geq4$, we have an operator of order $n$ in $\lambda$ of the form
$\nabla_4^{n_4}\nabla_5^{n_5}\nabla_6^{n_6}\nabla_7^{n_7}$ with
$n_4\geq0$, $n_5,n_6,n_7=0,1$ (at most one of them equal to 1). On the
other
hand, as every long derivative is of order$\geq4$, there are not operators
of order 1, 2, and 3. Then
\begin{eqnarray}\label{i3.1}
W_{\mbox{res}}(\bs)&=&
\mbox{span}\left\{\nabla_4^{n_4}\nabla_5^{n_5}\nabla_6^{n_6}\nabla_7^{n_7}
\chi(\bs,\lambda),
n_4\geq0,n_5,n_6,n_7=0,1\right.\nonumber\\
  &  &\quad\quad\quad\left.\mbox{at most one of }n_5,n_6,n_7\mbox{ equal
to }1\right\}
\end{eqnarray}
so that
$$S_{W_{\mbox{res}}(\bs)}={\Bbb N}-\{1,2,3\}$$
and consequently
$$\mbox{Index}(\dbar_{W_{\mbox{res}}(\bs)})=-3.$$
In order to get the linear problems involving the minimum number of
independent variables  we use, instead of (\ref{i3.1}) the more convenient
description of $W_{\mbox{res}}(\bs)$ as
\begin{equation}\label{i3.2}
W_{\mbox{res}}(\bs)=
\mbox{span}\{\nabla_4^{n_4}\nabla_5^{n_5}\chi,
n_4\geq0,n_5=0,1,2,3,\nabla_6\chi,\nabla_7\chi,
\nabla_5\nabla_6\chi \}.\end{equation}
The lowest order linear equations constructed using the system of
generators in (\ref{i3.2}) are
\begin{equation}\label{8.5}\begin{array}{l}
\partial_5^4\psi-\partial_4^5\psi+u_{19}\partial_5^3\partial_4\psi+
u_{18}\partial_5^2\partial_4^2\psi+
u_{17}\partial_4^3\partial_5\psi+u_{16}\partial_4^4\psi\\
    \\
\;\;+u_{15}\partial_5^3\psi+
u_{14}\partial_5^2\partial_4\psi+u_{13}\partial_5\partial_4^2\psi+
u_{12}\partial_4^3\psi+u_{11}\partial_5\partial_6\psi\\
    \\
\;\;+u_{10}\partial_5^2\psi+u_9\partial_4\partial_5\psi+u_8\partial_4^2\psi+
u_7\partial_7\psi+u_6\partial_6\psi\\
   \\
\;\;+u_5\partial_5\psi+u_4\partial_4\psi+u_0\psi=0,
\end{array}\end{equation}
\begin{equation}\label{8.6}\begin{array}{l}
\partial_4\partial_6\psi-\partial_5^2\psi+v_9\partial_4\partial_5\psi
+v_8\partial_4^2\psi+v_7\partial_7\psi+v_6\partial_6\psi\\
    \\
\;\;+v_5\partial_5\psi+v_4\partial_4\psi+v_0\psi=0,
\end{array}\end{equation}
\begin{equation}\label{8.7}\begin{array}{l}
\partial_4\partial_7\psi-\partial_5\partial_6\psi+p_{10}\partial_5^2\psi
+p_9\partial_4\partial_5\psi+p_8\partial_4^2\psi+p_7\partial_7\psi\\
    \\
\;\;+p_6\partial_6\psi+p_5\partial_5\psi+p_4\partial_4\psi+p_0\psi=0,
\end{array}\end{equation}
\begin{equation}\label{8.8}\begin{array}{l}
\partial_5\partial_7\psi-\partial_4^3\psi+q_{11}\partial_5\partial_6\psi
+q_{10}\partial_5^2\psi+q_9\partial_4\partial_5\psi+q_8\partial_4^2\psi\\
    \\
\;\;+q_7\partial_7\psi+q_6\partial_6\psi+q_5\partial_5\psi+q_4\partial_4\psi
+q_0\psi=0,
\end{array}\end{equation}
\begin{equation}\label{8.9}\begin{array}{l}
\partial_6^2\psi-\partial_4^3\psi+w_{11}\partial_5\partial_6\psi
+w_{10}\partial_5^2\psi+w_9\partial_4\partial_5\psi+w_8\partial_4^2\psi\\
   \\
\;\;+w_7\partial_7\psi+w_6\partial_6\psi+w_5\partial_5\psi+w_4\partial_4\psi
+w_0\psi=0,
\end{array}\end{equation}
where $u_i$, $i=0,4,5,\dots,19$; $v_j$, $j=0,4,5,\dots,9$;
$p_k$, $k=0,4,5,\dots,10$; $q_l$, $l=0,4,5,\dots,11$; and $w_r$,
$r=0,4,5,\dots,11$ are functions of $b_1$, $b_2$, $b_3$, $\chi_n$
$(n\geq1)$
and their
derivatives. Besides, we have three constraints  on the coefficients of
the wave functions for each equation (\ref{8.5})-(\ref{8.9}).

Again, among these five equations, there are only three independent
ones
modulo (\ref{8.5}). For instance, acting on (\ref{8.6}) by $\partial_5^2$
on (\ref{8.8}) by $\partial_4^2$ and using (\ref{8.5}), (\ref{8.6})
and (\ref{8.7})
one gets an identity. So (\ref{8.8}) and similarly (\ref{8.9}) are
satisfied due to equations (\ref{8.5})-(\ref{8.7}). These three equations
provides a four-dimensional  system (being the independent variables
$s_4$, $s_5$, $s_6$, $s_7$). However, using equation (\ref{8.6}) one can
get
$\partial_7\psi$ in terms of derivatives of $\psi$ with respect to the
three others
independent variables, i.e.
\begin{eqnarray}\label{8.13}
\partial_7\psi&=&-\frac{1}{v_7}(\partial_4\partial_6\psi-\partial_5^2\psi+
v_9\partial_4\partial_5\psi+v_8\partial_4^2\psi\nonumber\\
 & & \\
 &
&+v_6\partial_6\psi+v_5\partial_5\psi+v_4\partial_4\psi+v_0\psi),\nonumber
\end{eqnarray}
and then, this term can be eliminated from (\ref{8.5}) and (\ref{8.7}).
As a result one gets two linear equations for $\psi$ which contain only
$\partial_4$,
$\partial_5$ and $\partial_6$. The linear problem determined by this
couple of
equations is compatible by construction, and the compatibility condition
gives rise to a
system of nonlinear equations in three dimensions ($s_4$, $s_5$ and
$s_6$).
A new system in the hierarchy can be obtained as the compatibility
condition of one
of the equations considered before (where $\partial_7$ has been
eliminated)
and
(\ref{8.13}). In this case we have a system of 3+1-dimensional nonlinear
equations ( spatial variables $s_4$, $s_5$, $s_6$ and time $s_7$).
Linear problems containing higher times $s_8$, $s_9$,... have the form
\begin{equation}\label{8.14}
\partial_k\psi=\sum_{n_4\,n_5\,n_6}v_{n_4\,n_5\,n_6}\partial_4^{n_4}\partial_5^{n_5}
\partial_6^{n_6}\psi,\quad k=8,9,\dots,
\end{equation}
where $n_4,n_5\geq0$, $n_6=0,1$ and $v_{n_4\,n_5\,n_6}$ are certain
functions.
Note that in (\ref{8.14}) we have already eliminated the term
$\partial_7\psi$
by using (\ref{8.6}).

The compatibility conditions of (\ref{8.14}) with the equation obtained
by
eliminating
$\partial_7\psi$ in (\ref{8.5}) defines an infinite hierarchy of
3+1 dimensional nonlinear
systems with four independent variables: $s_4$, $s_5$, $s_6$, and time
$s_k$.
Now, one could eliminate also $\partial_6\psi$ from equations (\ref{8.5}),
(\ref{8.6}) and (\ref{8.7}), in order to get a single linear equation
containing
derivatives with respect to only two independent variables, $s_4$ and
$s_5$. But
in order to do it, one has to invert an involved differential operator.
Consequently,
the corresponding 2+1 dimensional equation is a complicated
integro-differential one.

Finally, from the above discusion, it is clear that we can eliminate
$\nabla_7\chi$
from (\ref{i3.2}), then we get a basis of $W_{\mbox{res}}(\bs)$ in the
form:
$$W_{\mbox{res}}(\bs)=
\mbox{span}\{\nabla_4^{n_4}\nabla_5^{n_5}\chi,
n_4\geq0,n_5=0,1,2,3,\nabla_6\chi,
\nabla_5\nabla_6\chi \}.$$

We finish the study of the hidden KP hierarchies by summarizing the
results
for the
general case $m\geq3$. Solving the equations (\ref{4.6}) with respect to
the first
$m$ times one has
$$t_k=b_k(s_{m+1},s_{m+2},\dots),\quad k=1,2,\dots,m,$$
then
$$g=\exp\left\{\sum_{k=1}^m\lambda^kb_k(\bs)+
\sum_{k=m+1}^{\infty}\lambda^ks_k\right\}$$
and
$$\nabla_k=\frac{\partial}{\partial s_k}+\sum_{l=1}^m\lambda^l
\frac{\partial b_l}{\partial s_k}+\lambda^k,\quad k=m+1,m+2,\dots.$$
In this case
$$S_{W_{\mbox{res}}(\bs)}={\Bbb N}-\{1,2,\dots,m\},$$
and consequently
$$\mbox{Index}(\dbar_{W_{\mbox{res}}(\bs)})=-m.$$
Now, by constructing operators of the form
$$L=\sum_{n_{m+1}\,n_{m+2},\dots}u_{n_{m+1}\,n_{m+2},\dots}
\nabla_{m+1}^{n_{m+1}}\nabla_{m+2}^{n_{m+2}}\cdots,$$
which satisfy the condition (\ref{2.15}), one gets an infinite hierarchy
of linear
equations. One can see that there are $m$ of them which form a basis of
Manakov ring of operators of lowest order with minimal number of
independent
variables $(s_{m+1},\;s_{m+2},\dots,s_{2m+1})$. As above, one can in
general eliminate
$\partial_{2m+1}\psi$, $\partial_{2m}\psi$,\dots,$\partial_{m+4}\psi$ from
these subsystem, and get a system of three-dimensional linear problems.
In this way, the whole hierarchy consists in 3+1 dimensional nonlinear
systems with constraints.

We finally point out that 3+1 hierarchies of integrable systems with
constraints have been discussed in a different situation  in
\cite{[32]}-\cite{s}.

\setcounter{equation}{0}

\section{Hidden Gelfand-Dikii hierarchies on the Grassmannian}

As a particular case of hidden KP hierarchies associated with sectors of
Gr
with nonzero $\dbar$ index, we discuss here the hidden Gelfand-Dikii (GD)
hierarchies, $1+1$ dimensional integrable hierarchies associated with
energy-dependent spectral problems. We prove that under certain
assumptions the only hidden
GD hierarchies are those associated with Schr\"{o}dinger equations with
energy-dependent potentials (hidden KdV hierarchies)
$$\partial_x^2\psi=\left(k^{2m+1}+\sum_{n=0}^{2m}u_n(\bs)k^n\right)\psi,
\quad k:=\lambda^2,$$
and that associated to the third-order equation (hidden Boussinesq
hierarchy)
$$\partial_x^3\psi=(k^2+u_1(\bs)k+u_0(\bs))\psi+(v_1(\bs)k+v_0(\bs))\partial_x\psi,
\quad k:=\lambda^3.$$

The hidden KdV hierarchies were already introduced and studied from the
point of view
of the Hamiltonian formalism in \cite{lma} and they were
 further generalized and
analyzed in \cite{af1}-\cite{rs}. As for the hidden Boussinesq hierarchy,
it is
one of the four cases studied in \cite{af2} in connection with the theory
of energy-dependent third-order Lax operators.
In both cases we manage to formulate a general solution method.

The start point here is a $l$-GD wave function. It is a particular KP wave
function
$\psi(\bt,\lambda)$ verifying the reduction conditions:
\begin{equation}\label{red}
\partial_{ml}\chi=0,\quad m=1,2,\dots.\end{equation}
As a consequence, its corresponding flow can be characterized as
$$W(\bt)=\mbox{span}\{\lambda^{ml}\nabla_1^n\chi(\bt,\lambda),m\geq0,
0\leq n\leq l-1\}.$$
Note that $W(\bt)$ does not depend on the parameters $t_{ml}$
$m=1,2,\dots$, so from now on
they will be supposed to be set equal zero, or equivalently
$$\bt=(t_1,\dots,t_n,\dots),\quad n\not\in(l):=\{l,2l,3l,\dots\}.$$
In the $\dbar$-dressing method, reductions (\ref{red}) correspond
to kernels of the form
$$R_0(\mu,\bar{\mu},\lambda,\bar{\lambda})=\delta(\mu^l-\lambda^l)\tilde{R}_0
(\mu),$$
then, we have the $\dbar$-problem
\begin{equation}\label{dbgd}
\frac{\partial\chi(\bt,\lambda)}{\partial\bar{\lambda}}=
\sum_{\alpha=1}^l\chi(\bt,q^{\alpha}\lambda)\tilde{R}_{\alpha}(\bt,\lambda)
\end{equation}
where
$$\begin{array}{l}
q=\exp\left(\frac{2\pi i}{l}\right),\\
   \\
\tilde{R}_{\alpha}(\bt,\lambda)=g_0(\bt,\lambda)\tilde{R}_{\alpha}(\lambda)
g_0^{-1}(\bt,q^{\alpha}\lambda)\end{array}$$
with $g_0(\bt,\lambda)=\exp\left(\sum_{i\not\in(l)}t_i\lambda^i\right)$
and $\tilde{R}_{\alpha}(\lambda)$ $\alpha=1,2,\dots,l$ are $l$ arbitrary
functions.
Note that the $\dbar$ problem (\ref{dbgd}) is invariant under
multiplication
by $\lambda^l$.

\vspace{3mm}

We focus now our attention in the study of the hidden l-GD hierarchy.
In order to do that, given an integer number $r>0$, $r\not\in(l)$, we
consider restriction under
$d_r=r-1-[\frac{r-1}{l}]$ constraints (\ref{4.6}), (here $[\cdot]$ denotes
integer
part),
solved with respect to the first consecutive $d_r$ parameters. It means
$$t_i=b_i(\bs_r),\quad 1\leq i<r,\quad i\not\in(l)$$
being $\bs_r=(s_r,\dots,s_n,\dots)$ $n>r$, $n\not\in(l)$. We have then
that
\begin{eqnarray}\label{9.1}
W_{\mbox{res}}(\bs_r)&=&\mbox{span}\{\lambda^{ml}
\nabla_{i_1}^{n_1}\nabla_{i_2}^{n_2}\cdots\chi(\bs_r,\lambda),
m,n_1,n_2,\dots\geq 0,
i_1,i_2,\dots\geq r,\nonumber\\
 & &\quad\quad\quad i_1,i_2,\dots\not\in(l)\}
\end{eqnarray}
and long derivatives are here defined as
$$\nabla_n=\frac{\partial}{\partial s_n}+\sum_{i<r}\lambda^i
\frac{\partial b_i}{\partial s_n}+\lambda^n, \quad n\geq r.$$
Clearly, if we look for $(1+1)$-dimensional hierarchies associated to
these
restrictions we need
\begin{equation}\label{9.2}
W_{\mbox{res}}({\bs_r})=\mbox{span}\{\lambda^{ml}\nabla_x^n\chi(\bs_r,\lambda),
m\geq0,
0\leq n\leq l-1\}
\end{equation}
where $x$ stands now for $s_r$. Consequently we are interested in those
submanifolds
${\cal M}$ for which the corresponding $W_{\mbox{res}}({\bs_r})$  verifies
(\ref{9.2}). In this sense we have:

\begin{pro} The family $W_{\mbox{res}}({\bs_r})$ satisfies (\ref{9.2}) if
and only if
the function $\psi(\bs_r,\lambda)$
obeys an infinite system of linear equations of the form
\begin{equation}\label{9.3}\begin{array}{l}
\partial_x^l \psi=\sum_{m=0}^{l-1}u_m (\bs_r,k)
\partial_x^m \psi,
\\
\\
\partial_n\psi=\sum_{m=0}^{l-1}\alpha_{nm} (\bs_r,k)
\partial_x^m
\psi,\quad   n> r,\;\; n\not\in (l)
\end{array}
\end{equation}
where $u_m$ and $\alpha_{nm}$ are polynomials in $k:=\lambda^l$.
\end{pro}
{\bf Proof:} The function $\chi(\bs_r,\lambda)$ as well as  its long
derivatives of all orders with
respect to  the variables $s_n$ belong to $W_{\mbox{res}}({\bs_r})$.
Then, if (\ref{9.2}) holds all these functions can be  decomposed in terms
of the basis
 $k^m\nabla_x^n\chi$. Therefore
(\ref{9.3}) follows.

Reciprocally, if $\chi$ satisfies a system of the form
(\ref{9.3}) then from (\ref{9.1}) and by using  Taylor expansion
we deduce (\ref{9.2}) at once.$\Box$

The next statement describes the cases
in which (\ref{9.2}) may arise.

\begin{pro}
Only two classes of parametrized
submanifolds ${\cal M}$ satisfying (\ref{9.2}) are allowed:
\begin{description}
\item[i)] Submanifolds ${\cal M}^{(2)}_{m}$of the form
$$
t_{2i-1}=b_i(\bs_{2m+1}),\; i=1,\ldots,m,\; \;\;m\geq 1,
$$
for the $2^{nd}$ GD hierarchy.
\item[ii)] Submanifolds ${\cal M}^{(3)}$ of the form
$$
t_1=b(\bs_2),
$$
for the $3^{rd}$ GD hierarchy.
\end{description}
\end{pro}
{\bf Proof:} Let us assume that (\ref{9.2}) holds
then $W_{\mbox{res}}({\bs_r})$ has no
elements with order $n$ such that $0<n<\text{min}(r,l)$. Moreover
$$
\mbox{order}(\lambda^{lm}\nabla_x^n\chi)=lm+nr.
$$
Let us first consider the case $r>l$.  It implies that $l+r<2r$, and
therefore if there are $i\not\in (l)$ such that
 $r<i<l+r$ then the functions $\nabla_i\chi$
are elements of order $i$ which cannot be decomposed in terms
of the basis $\lambda^{lm}\nabla_x^n\chi$.
The only
way to avoid these functions is to take
$l=2$, and consequently the allowed $r$ are the odd integers $r=2m+1$
$(m\geq 1)$.

Consider now the case $r<l$. We have
$$
\mbox{order}(\nabla_x\chi)<\mbox{order}(\nabla_x^2\chi)
<\mbox{order}(\lambda^{l}\nabla_x\chi).
$$
Thus, given  $i\not \in (l)$ such that
 $r<i<2r$ then the functions  $\nabla_i\chi$
are elements of order $i$ which cannot be decomposed in terms of the basis
$\lambda^{lm}\nabla_x^n\chi$. But it is obvious that these functions will
arise unless we take $r=2$ and
 $l=3$.$\Box$

\vspace*{.5cm}

As it will be proved below
 one can construct explicit examples of submanifolds ${\cal M}^{(2)}_m$
and ${\cal M}^{(3)}$ satisfying (\ref{9.2}). Observe that in these
cases the corresponding families of subspaces in the Grassmannian lead to
the following values of the index of $\dbar$:
\begin{description}
\item[1)]For ${\cal M}^{(2)}_m$:
$$
S_{W_{\mbox{res}}({\bs_{2m+1}}})={\Bbb N}-\{1,3,\ldots,2m-1\},
$$
so that
$$
\mbox{Index}(\dbar_{W_{\mbox{res}}({\bs_{2m+1}}}))=-m.
$$
\item[2)]For ${\cal M}^{(3)}$:
$$
S_{W_{\mbox{res}}({\bs_2})}={\Bbb N}-\{1\},
$$
and as a consequence
$$
\mbox{Index}(\dbar_{W_{\mbox{res}}({\bs_2})})=-1.
$$
\end{description}
In both cases the families of subspaces in the Grassmannian lie outside
the zero index sector of the $\dbar$ operator. Next,
we are going to show that for  both classes of submanifolds
described above there exist hierarchies of integrable systems.

Before analyzing these two cases, it is worth noticing that we have only
looked
for submanifolds associated to $(1+1)$-dimensional hierarchies of
integrable systems, obtained by solving constraints (\ref{4.6}) with
respect
to the first variables. Nevertheless, by using the methods of the previous
section
and solving the constraints with respect to any set of variables, we can
get in
general multidimensional hidden $l$-Gelfand-Dikii hierarchies for
arbitrary $l$ and $r$
$(r\not\in(l))$. It is also clear that all these hierarchies belong
to sectors in the Grassmannian with nonzero index.

\subsection{Hidden KdV hierarchies}

 Consider first submanifolds ${\cal M}_m^{2}$ verifying (\ref{9.2}).
>From Proposition 1  the constrained wave function
$\psi(\bs_{2m+1},\lambda)$
satisfies an  infinite linear system  of the form
\begin{equation}\label{9.4}\begin{array}{l}
\partial_x^2 \psi =u(\bs_{2m+1},k)\psi,
\\
\\
\partial_{2n+1} \psi =\alpha_n(\bs_{2m+1},k)\psi +
\beta_n( \bs_{2m+1},k)\partial_x\psi,\quad n>m,
\end{array}
\end{equation}
where $k:=\lambda^2$, $u=u(\bs_{2m+1},k):=k^{2m+1}+\sum_{n=0}^{2m}k^n
u_n(\bs_{2m+1})$,
and $\alpha_n$, $\beta_n$ are polynomials in $k$.
By introducing the bilinear form
$$
B(\psi,\varphi):=-\frac{1}{2\lambda^{2m+1}}\left| \begin{array}{cc}
\psi(\lambda)&\varphi(\lambda)\\
\psi(-\lambda)&\varphi(-\lambda)
\end{array}
\right |,
$$
we may write the coefficients $\alpha_n$ and $\beta_n$ in (\ref{9.4}) as
\begin{equation}\label{9.5}
\alpha_n=\frac{B(\partial_{ 2n+1}\psi ,\partial_x \psi )}
{B(\psi ,\partial_x \psi )},\quad
\beta_n=\frac{B(\psi ,\partial_{ 2n+1}\psi )}
{B(\psi ,\partial_x \psi )}.
\end{equation}
Then, we have $\alpha_n=-\frac{1}{2}\beta_{nx}$.
Furthermore, the compatibility conditions for (\ref{9.4})
imply
\begin{equation}\label{9.6}
\partial_{ 2n+1}u=J\beta_n,\quad \mbox{where} \quad
J:=-\frac{1}{2}\partial_x^3+2u\partial_x+u_x.
\end{equation}

On the other hand, the function $\beta_n$ is related to the trace
of the resolvent of the Schr\"odinger operator
$$
R( \bs_{2m+1},k):=\frac{\psi(\bs_{2m+1},\lambda)\psi(\bs_{2m+1},-\lambda)}
{B(\psi ,\partial_x \psi )}.
$$
Thus, from (\ref{9.5}) and the polynomial character of $\beta_n$ as a
function of $k$ it follows that
$$\beta_n=(k^{n-m}R)_+,$$
where $(k^{n-m}R)_+$ stands for the polynomial part of $k^{n-m}R$ with
$R$ being substituted
 by its expansion as $k\rightarrow\infty$
$$
R=1+\sum_{n\geq 1}\frac{R_n(\bs_{2m+1})}{k^n}.
$$
Therefore
\begin{equation}\label{9.7}
\partial_{ 2n+1}u=J(k^{n-m}R)_+.
\end{equation}
It turns out \cite{lma} that the coefficients $R_n$ are differential
polynomials
in the potential functions $(u_0,u_1,\ldots,u_{2m})$. They can be
determined by identifying coefficients of powers of $k$ in the equation
$$
JR=0.
$$
In this way the set of equations  (\ref{9.7}) constitutes a
hierarchy of integrable systems associated with the Schr\"odinger operator
in
(\ref{9.4}). We will refer to this hierarchy as $\mbox{KdV}_{2m+1}$.
Solutions of  the  members of the hierarchies can be derived from the
functions
$b_i$ and $\chi_n$.

For example, for $m=1$, using our standard techniques and the method
considered above to construct the hierarchy, we have that the potential
function is given by
\begin{equation}\label{9.8}
\begin{array}{l}
u_2=2b_x,\\
u_1=b_x^2+2\chi_{1x}, \\
u_0=2\chi_{3x}-2\chi_2\chi_{1x} +
\chi_1 b_{zz} + 2\chi_{1x}b_x,
\end{array}
\end{equation}
where $b:=b_1$. The first integrable system in the hierarchy corresponds
to
$t:=s_5$ and takes the form
\begin{equation}\label{9.10}
\begin{aligned}
\partial_t u_0&=\frac{1}{4} u_{2xxx}-u_0 u_{2x}-\frac{1}{2}
u_2 u_{0x},\\
\partial_t u_1&=-\frac{1}{2}u_2 u_{1x}-
u_1  u_{2x}+\  u_{0x},\\
\partial_t u_2&=-\frac{3}{2}
u_2 u_{2x}+  u_{1x}.
\end{aligned}
\end{equation}
The second equation in (\ref{9.4}) is in this case
$$\partial_t\psi=\frac{1}{4}u_{2x}\psi+\left(k-\frac{1}{2}u_2\right)\partial_x\psi,$$
and the absence of the pure first order $\lambda$ in
$W_{\mbox{res}}({\bs_2})$
means
(eq.(\ref{9.4})) the constraints
\begin{equation}\label{c1}
\begin{array}{l}
\chi_{1x}=b_x^2+b_t\\
\chi_{2x}=\chi\chi_{1x}-\frac{1}{2}b_{xx}.
\end{array}\end{equation}
Note also, that using (\ref{9.8}) and the first equation in (\ref{c1}),
system (\ref{9.10}) is equivalent to a system of two equations for $b$ and
$u_0$
$$\begin{array}{l}
u_{0t}=\frac{1}{2}b_{xxxx}-2b_{xx}u_0-b_xu_{0x},\\
   \\
(2b_t+3b_x^2)_t+2b_{xx}(2b_t+3b_x^2)-u_{0x}+b_x(2b_t+3b_x^2)_x=0.
\end{array}$$

Analogously, for $m=2$, we have that the potential function is given by
$$\begin{aligned}
u_4=&2  b_{2x}\\
u_3=&2 b_{1x}+ b_{2x}^2,\\
u_2=&2  b_{1x}  b_{2x}+2\chi_{1x},\\
u_1=&2 \chi_{3x}-2\chi_2  \chi_{1x}+
\chi_1 b_{2xx} +
+2 \chi_{1x} b_{2x}+ b_{1x}^2, \\
u_0=&2 \chi_{5x}-2\chi_4  \chi_{1x}+\chi_3 b_{2xx}+2\chi_{3x} b_{2x}
+\chi_1  b_{1xx}\\ &
+ 2 b_{1x} \chi_{1x}-(u_1-b_{1x}^2) \chi_{2x}.
\end{aligned}$$
The second equation in (\ref{9.4}) is
$$\partial_t\psi=\frac{1}{4}u_{4x}\psi+\left(k-\frac{1}{2}u_4\right)
\partial_x\psi,$$
and the absence of orders $\lambda$ and $\lambda^3$ means the constraints
$$\begin{array}{l}
b_{2x}^2+b_{2t}-b_{1x}=0,\\
\chi_{1x}-b_{2x}b_{1x}-b_{1t}=0,\\
2\chi_{2x}+b_{2xx}-2\chi_{1x}\chi_1=0,\\
2\chi_{4x}+b_{1xx}-2(\chi_1\chi_3)_x+b_{2xx}(\chi_2-\chi_1^2-b_{2x})+2\chi_2\chi_1\chi_{1x}=0.
\end{array}$$

Finally, we point out that although we have only considered constraints
solved
with respect to the first parameters, submanifolds of
the form ${\cal M}_m^{(2)}$ can be  parametrized in other ways and we also
get
hierarchies of integrable systems. Unfortunately, in this case , there are
not avaible direct methods to construct the hierarchies, as the one
discused
above to construct the KdV$_{2m+1}$ hierarchy, but we can always use
standard
techniques to get integrable systems. For example, taking the case $m=1$
with the parametrization
$$t_3=b(\bs),\quad \bs:=(s_1,s_5,\dots,s_{2m+1},\dots)$$
and taking $x:=s_1$, $t:=s_5$ we have the linear problem
\begin{equation}\label{slp}\begin{array}{l}
\partial_x^2\psi=(u_3k^3+u_2k^2+u_1k+u_0)\psi+v_0\partial_x\psi,\\
\partial_t\psi=\alpha_0\psi+(\beta_1k+\beta_0)\partial_x\psi\end{array}\end{equation}
where
\begin{equation}\label{sp}\begin{array}{l}
u_3=b_x^2,\quad u_2=2b_x,\quad u_1=1+2b_x\chi_{1x},\\
    \\
u_0=-\frac{b_{xx}}{b_x}\chi_1-2b_x\chi_{1x}\chi_2+2\chi_{1x}+2b_x\chi_{3x},\\
   \\
v_0=\frac{b_{xx}}{b_x},\quad \alpha_0=-2\frac{u_{2x}}{u_2^3},\quad
\beta_1=\frac{2}{u_2},\quad \beta_0=2\frac{1-u_1}{u_2^2},
\end{array}\end{equation}
and the absence of order $\lambda$ in equation (\ref{slp}) means the
constraints
\begin{equation}\label{sc}\begin{array}{l}
\chi_{1x}=\frac{1}{b_x}-b_t,\\
   \\
\chi_{2x}=\chi\chi_{1x}+\frac{b_{xx}}{2b_x^2}.
\end{array}\end{equation}
By imposing the compatibility condition in the linear problem (\ref{slp})
we have the integrable system
$$\begin{array}{l}
u_{3t}=2\beta_{1x}u_2+2\beta_{0x}u_3+\beta_1u_{2x}+\beta_0u_{3x},\\
u_{2t}=2\beta_{1x}u_1+2\beta_{0x}u_2+\beta_1u_{1x}+\beta_0u_{2x},\\
u_{1t}=2\beta_{1x}u_0+2\beta_{0x}u_1+\beta_1u_{0x}+\beta_0u_{1x},\\
u_{1t}=\alpha_{0xx}+2\beta_{0x}u_0+\beta_0u_{0x}-\alpha_{0x}v_0,\\
u_{0t}=(2\alpha_0+\beta_{0x}+v_0\beta_0)_x,
\end{array}$$
that using (\ref{sp}) and (\ref{sc}) can be reduced to a system of
two equations for $b$ and $u_0$:
$$\begin{array}{l}
(b_xb_t)_t-\frac{b_{xx}}{b_x^2}u_0+\left(\frac{b_t}{b_x}-\frac{1}{b_x^2}\right)_x
(3-2b_xb_t)+\frac{1}{2b_x}u_{0x}-\left(\frac{b_t}{b_x}-\frac{1}{b_x^2}\right)
b_xb_t=0,\\
   \\
u_{0t}+\frac{1}{2}\left(\frac{b_{xx}}{b_x^3}\right)_{xx}-2
\left(\frac{b_t}{b_x}-\frac{1}{b_x^2}\right)_xu_0-
\left(\frac{b_t}{b_x}-\frac{1}{b_x^2}\right)u_{0x}-
\frac{b_{xx}}{2b_x }\left(\frac{b_{xx}}{b_x^3}\right)_x=0.
\end{array}$$

\subsection{Hidden Boussinesq hierarchy}

Our next task is to show that the submanifolds ${\cal M}^{(3)}$
satisfying
(\ref{9.2})  are also associated with a hierarchy of integrable systems.
>From Proposition 1, now, we have
\begin{equation}\label{9.11}\begin{array}{l}
\partial_x^3 \psi =u( \bs_2,k)\psi + v( \bs_2,k)\partial_x\psi,
 \\
\partial_n \psi =\alpha_n( \bs_2,k)\psi_ +\beta_n( \bs_2,k)
\partial_x\psi_ +\gamma_n( \bs_2,k)\partial_x^2\psi ,
\end{array}
\end{equation}
where  $k:=\lambda^3$, $u:=u_0(\bs_2)+k u_1(\bs_2)+k^2$,
$v:=v_0(\bs_2)+k v_1(\bs_2)$
and $\alpha_n$, $\beta_n$ and $\gamma_n$ are polynomials in $k$.
By introducing the trilinear form
$$
T(\psi,\varphi,\eta):=\frac{1}{3i\sqrt{3}\lambda^6}
\left| \begin{array}{ccc}
\psi(\lambda)&\varphi(\lambda)&\eta(\lambda)\\
\psi(\epsilon \lambda)&\varphi(\epsilon \lambda)&\eta(\epsilon \lambda)\\
\psi(\epsilon^2 \lambda)&\varphi(\epsilon^2 \lambda)&\eta(\epsilon^2
\lambda)
\end{array}
\right |,
$$
where $\epsilon=(-1+i\sqrt{3})/2$, we may write the coefficients in
(\ref{9.11}) as
$$\begin{array}{c}
\alpha_n=
\displaystyle\frac{T(\partial_{n}\psi,\partial_x \psi,\partial_x^2
\psi)}
{T(\psi,\partial_x \psi,\partial_x^2\psi)},\quad
\beta_n=\displaystyle\frac{T(\psi,\partial_{n}\psi,\partial_x^2 \psi)}
{T(\psi,\partial_x \psi,\partial_x^2\psi)},\\ \\
\gamma_n=\displaystyle\frac{T(\psi,\partial_x \psi
,\partial_{n} \psi )}
{T(\psi ,\partial_x \psi ,\partial_x^2\psi )}.
\end{array}
$$
By using now (\ref{9.11}) it immediately follows that
$$
\alpha_n=-\frac{1}{3}\left [ 2v \gamma_n+\gamma_{nxx}+3\beta_{nx}\right ].
$$
Futhermore, the compatibility conditions for (\ref{9.11})
imply
$$
\left( \begin{array}{c}
\partial_n v\\
\partial_n u
\end{array}
\right )=
J
\left( \begin{array}{c}
\gamma_{nx} +\beta_n\\
\gamma_n
\end{array}
\right ),
$$
where $J$ is the matrix operator given by
$$
\begin{array}{c}
J_{11}:=-2\partial_x^3+2v\partial_x+ v_x,\;\;
J_{12}:=\partial_x^4-\partial_x^2\cdot v+3\partial_x \cdot u-u_x,\\ \\
J_{21}:=-\partial_x^4+v\partial_x^2+3u\partial_x+u_x,\\ \\
J_{22}:=\frac{1}{3}[ 2\partial_x^5-2(v\partial_x^3+\partial_x^3\cdot v)+
(v^2+3u_x)\partial_x+\partial_x\cdot (v^2+3u_x)].
\end{array}
$$
The standard technique shows that these equations
are related with the resolvent trace functions
$$
R( \bs_2,k):=
\left (
\begin{array}{c}
\displaystyle\frac{T(\psi ,\partial_x \psi,\lambda\partial_x \psi)}
{T(\psi,\partial_x \psi,\partial_x^2\psi)}\\ \\
\displaystyle\frac{T(\psi,\partial_x \psi,\lambda\psi)}
{T(\psi,\partial_x \psi,\partial_x^2\psi)}
\end{array}
\right ),\;\;
S( \bs_2,k):=
\left (
\begin{array}{c}
\displaystyle\frac{T(\psi,\partial_x \psi,\lambda^2\partial_x \psi)}
{T(\psi,\partial_x \psi,\partial_x^2\psi)}\\ \\
\displaystyle\frac{T(\psi,\partial_x \psi,\lambda^2\psi)}
{T(\psi,\partial_x \psi,\partial_x^2\psi)}
\end{array}
\right ),
$$
in the form
\begin{equation}\label{9.12}
\left( \begin{array}{c}
\partial_{ 3n+1} v\\
\partial_{ 3n+1} u
\end{array}
\right )=
J (k^n R)_+,\;\;
\left( \begin{array}{c}
\partial_{ 3n-1} v\\
\partial_{ 3n-1} u
\end{array}
\right )=
J (k^{n-1} S)_+.
\end{equation}
In these expressions  $R$ and $S$ are substituted by their
expansion as $k\rightarrow\infty$
$$
R=\sum_{n\geq 1}\frac{R_n(\bs_2)}{k^n},\quad \mbox{and}\quad
S=\sum_{n\geq 0}\frac{S_n(\bs_2)}{k^n},.$$
The coefficients $R_n$ are $S_n$ are differential polynomials
in the potential functions $u_i,v_i$ $(i=0,1)$. They can be
determined by identifying coefficients of powers of $k$ in the equations
$$
JR=0,\;\;\; JS=0.
$$
It is easy to prove that Eq. (\ref{9.12}) constitutes an evolution
equation
for $u$ and $v$. The set of these equations is a
 hierarchy of integrable systems
associated with the third-order operator in (\ref{9.11}).
We will refer to this hierarchy as the hidden Boussinesq hierarchy.
Solutions of  the  members of the hierarchy can be derived from the
functions
$b$ and $\chi_n$.
\begin{equation}\label{9.13}
\begin{array}{ll}
u_0=&b_{xxx}\chi_1-3\chi_{1x}\chi_3+3\chi_{4x}+3b_{x}\chi_{3x}+
3b_{x}^2\chi_{2x}+\\ \\&
+3b_{xx}\chi_{1x}+3\chi_{2xx}+3b_{x}\chi_{1xx}+3\chi_{1x}\chi_1\chi_2-
3\chi_{2x}\chi_2-\\ \\&
-3b_{x}\chi_{1x}\chi_2-3b_{xx}b_{x}^2\chi_1+3\chi_{1x}\chi_1^2b_{x}-
3\chi_{2x}\chi_1b_{x}-3b_{x}^2\chi_{1x}\chi_1,
\\\\
u_1=&b_{x}^3+3b_{xx}+3\chi_{1x},\\ \\
v_0=&3b_{xx}b_{x}-3\chi_{1x}\chi_1+3\chi_{2x}+3b_{x}\chi_{1x},\\ \\
v_1=&3b_{x}.
\end{array}
\end{equation}

The first equation of the hidden Boussinesq hierarchy corresponds
to the time parameter $t:=s_4$ and takes the form
\begin{equation}\label{9.14}
\begin{array}{ll}
\partial_t u_0=&\frac{2}{9}v_1v_{1xxxx}+
\frac{8}{9}v_{1x}v_{1xxx}
+\frac{2}{3}v_{1xx}^2-
\frac{2}{9}v_0v_1v_{1xx}-
\frac{2}{9}v_0v_{1x}^2-
\frac{2}{3}u_0v_1v_{1x}-\\ \\ &
\frac{1}{9}u_{0x}v_1^2-
\frac{2}{3}v_{0xxx}+
\frac{2}{3}v_0v_{0x}+u_{0xx},\\ \\
\partial_t u_1=&-\frac{2}{9}v_1^2v_{1xx}-
\frac{2}{9}v_1v_{1x}^2-
\frac{2}{3}u_1v_1v_{1x}-
\frac{1}{9}u_{1x}v_1^2-
\frac{2}{3}v_{1xxx}+
\frac{2}{3}(v_0v_1)_x+u_{1xx},\\ \\
\partial_t v_0=&\frac{4}{9}v_1v_{1xxx}+
\frac{4}{3}v_{1x}v_{1xx}-
\frac{4}{9}v_0v_1v_{1x}-
\frac{1}{9}v_1^2v_{0x}-v_{0xx}+2u_{0x},\\ \\
\partial_t v_1=&-\frac{5}{9}v_1^2v_{1x}-v_{1xx}+2u_{1x},
\end{array}
\end{equation}
The second equation in the linear problem (\ref{9.11}) is
$$\partial_t\psi=(\alpha_0+\alpha_1k)\psi+\beta_0\partial_x\psi+
\partial_x^2\psi$$
with
$$\alpha_0=\frac{2}{9}v_1v_{1x}-\frac{2}{3}v_0,\quad
\alpha_1=-\frac{2}{3}v_1,\quad \beta_0=-\frac{1}{9}v_1^2.$$
Finally, the constraints imposed by the absence of the first order in
$W_{\mbox{res}}({\bs_2})$ are given by
\begin{equation}\label{9.15} \begin{array}{c}
\frac{1}{3}b_{xxx}-\chi_{1x}\chi_2+\chi_{3x}+\chi_{1xx}+\chi_{1x}\chi_1^2-
 \chi_{2x}\chi_1
-b_{xx}b_x^2=0,\\
 b_x^3+b_t-b_{xx}-2\chi_{1x}=0.\end{array}\end{equation}

\subsection{Methods of Solution of hidden Gelfand-Dikii Hierarchies}

A solution method for the hierarchies studied above was discussed for the
hidden KdV hierarchies in \cite{km} and some solutions were exhibited
there.
The main idea is that taking a particular $l$-GD wave function ($l=2$ for
${\cal M}^{(2)}_m$ and $l=3$ for ${\cal M}^{(3)}$) the constraints imposed
by
the absence of some orders in $W_{\mbox{res}}({\bs_r})$
determine differential equations for the functions $b_i$ associated
with the submanifolds ${\cal M}^{(2)}_m$ and ${\cal M}^{(3)}$ satisfying
(\ref{9.2}). If these equations can be solved, they lead to solutions
of the corresponding hierarchy. In general, these differential equations
are too complicated to be solved.  Nevertheless, we may provide
appropriate methods
of solution directly based on the Grassmannian.  To this end it is
required
an element $W$ of $\gr$ associated to a
wave function $\psi$ for the $l$-Gelfand-Dikii hierarchy, such that the
functions of $W$ admit
 meromorphic expansions in the disk $D_0={\Bbb C}-D_{\infty}$ with fixed
 poles $\lambda_i,\; i=1,\ldots,n$
of maximal orders $r_i,\; i=1,\ldots,n$. Under these conditions any linear
functional on $W$ of the form
$$
l(w)=\sum_{j=1}^{s}c_j\frac{d^{n_j} w}{d \lambda^{n_j}}(q_j),
\quad |q_j|<1,\quad q_j\not\in \{\lambda_i:i=1,\ldots,n\},
$$
admits a representation
\begin{equation}\label{9.16}
l(w)=\oint_{S_{\infty}}\tilde{w}(\lambda)w(\lambda)\frac{d\lambda}{2\pi i
\lambda},
\quad \forall w\in W,
\end{equation}
with a finite order function $\tilde{w}$. For example in the case
$r_i=1,\,\forall i=1,\ldots,n$ we have
$$
\frac{1}{n!}\frac{{\rm d}^n w}{{\rm d}\lambda^n}(q)=\oint_{S_{\infty}}
\left[
\frac{\lambda}{(\lambda-q)^{n+1}}-\sum_{j}\frac{\lambda}{(\lambda_j-q)^{n+1}}
\prod_{i\not=j}
\frac{\lambda-\lambda_i}{\lambda_j-\lambda_i}\right] w(k)
\frac{d\lambda}{2\pi i \lambda}.
$$
\begin{pro} Given  $W\in\gr$  associated to a wave function
$\psi$ for the KdV hierarchy and  a linearly independent
 set of $m$ functionals $\{l_i\}_{i=1}^m$ such that for certain
numbers $c_{ij}$
\begin{equation}\label{9.17}
W\subset\mbox{Ker}\big(\lambda^2 l_i-\sum_j c_{ij} l_j\big ),\quad
 i=1,\ldots,m,
\end{equation}
where $(\lambda^2l_i)(w)\equiv l_i(\lambda^2 w)$. Then, a submanifold
 ${\cal M}_m^{(2)}$ satisfies (\ref{9.2}) if
\begin{equation}\label{9.18}
l_i(\psi(\bt,\lambda))=0,\; i=1,\ldots,m,
\end{equation}
for all $\bt\in{\cal M}_m^{(2)}$.
\end{pro}
{\bf Proof:} Let $\{\tilde{w_i}\; :\; i=1,\ldots,m\}$ be the functions
representing the functionals $l_i$ . From (\ref{9.17}) it follows that
$$
\lambda^2\tilde{w}_i=\sum_j c_{ij}\tilde{w}_j+\tilde{u}_i,
\quad i=1,\ldots,m,
$$
where $\tilde{u}_i$ are elements of $\widetilde{W}$. Moreover, from
(\ref{9.16}) we have
that the equations (\ref{9.18})
are equivalent to
$$
\oint_{S_{\infty}}\tilde{w}_i(\lambda)\psi(\bt,\lambda)
\frac{d\lambda}{2\pi i \lambda}=0,\quad i=1,\ldots,m.
$$
Therefore, if $t_{2i-1}=b_i(\bs_{2m+1})$, $i=1,\ldots,m$ are the
functions  characterizing  the parametrized submanifold ${\cal
M}_m^{(2)}$,
then,
 the restricted wave function $\psi(\bs_{2m+1},\lambda)$ generates
a subspace $W_{\mbox{res}}$ such that
$$
\widetilde{W_{\mbox{res}}}=\widetilde{W}\oplus \mbox{span}\{\tilde{w}_i,\;
i=1,\ldots,m\}.
$$
As a consequence
$$
\mbox{v.d.}(W_{\mbox{res}})=\mbox{v.d.}(W)-m=-m.
$$
Moreover, it is known \cite{ps} that the virtual dimension of a subspace
$W$ does
not change under the action of an invertible  multiplication operator.
Then, by taking (\ref{ind}) into account
we have
$$
\mbox{Index}\;\bar{\partial}_{W_{\mbox{res}}({\bs_{2m+1}})}
=\mbox{v.d.}(W_{\mbox{res}}({\bs_{2m+1}}))
=\mbox{v.d.}(W_{\mbox{res}})=-m .
$$
Hence, it is easy to deduce that
$$
S_{W_{\mbox{res}}({\bs_{2m+1}})}={{\Bbb N}}-\{1,3,\ldots,2m-1\},
$$
so that the statement follows at once.$\Box$

In the same way one proves:

\begin{pro} Given $W$ associated  to a wave function
 $\psi$  for the Boussinesq hierarchy
 and a non-trivial functional $l$ on $W$ verifying
$$
W\subset \mbox{Ker}\big(k^3 l-cl\big),
$$
for a certain number $c$. Then a submanifold ${\cal M}^{(3)}$ satisfies
(\ref{9.2}) if
\begin{equation}\label{9.19}
l(\psi( \bt,\lambda))=0,
\end{equation}
for all $\bt\in{\cal M}^{(3)}$.
\end{pro}

These results allow us to determine the functions $b_i=b_i(\bs_r)$ from
 constraints of the types (\ref{9.18})-(\ref{9.19}).

 \vspace{5mm}

We devote the rest of the section to illustrate this method by
constructing
some solutions. We concentrate first on the hidden KdV hierarchies.
Our first example
is based on the
subspace $W\in\gr$ of boundary values of functions $w=w(\lambda)$
analytic on the
unit disk $|\lambda|<1$ , with the
possible exception of a single real pole  $-1<q<1$ and such that
$$
\lambda^2W\subset W \quad \mbox{and}\quad Res(w,q)=cw(-q),
$$
for a given $c>0$. This subspace determines a KdV wave function
$$
\psi(\bt,\lambda)=\exp(\sum_{n\geq1}\lambda^{2n-1}t_{2n-1})
\left(1+\frac{a(\bt)}{\lambda-q}\right),\quad
a(\bt)=\frac{2qc(\bt)}{2q+c(\bt)},
$$
where
$$
c(\bt):=c\exp\left(-2\sum_{n\geq1}q^{2n-1}t_{2n-1}\right).
$$
We may construct solutions of the $\mbox{KdV}_{3}$ hierarchy from $W$ by
means of the functional
$$
l(w)=\frac{{\rm d} w}{{\rm d} \lambda}(0),
$$
which obviously satisfies $\lambda^2l=0$. The implicit equation
$l(\psi( \bt,\lambda))=0$
reads
$$
t_1(1-\frac{a(\bt)}{q})=\frac{a(\bt)}{q^2}.
$$
By introducing the new variables
$$\begin{array}{l}
y:=2q t_1,\; \; x:=s_3,\\ \\
z:=2q^3 x-\log c+2\sum_{n\geq 2} q^{2n+1}s_{2n+1},
\end{array}$$
the equation reduces to
\begin{equation}\label{9.20}
y+z=\log\left[\frac{1}{2q}+\frac{2}{qy}\right].
\end{equation}
For $q>0$ it defines  two branches $y^{(i)}=2qb^{(i)}(z)$ ($i=1,2$),
 while for $q<0$ it leads to only one
branch $y^{(3)}=2qb^{(3)}(z)$. Moreover, from (\ref{9.20}) we have
$$
\frac{{\rm d} y}{{\rm d} z}=-1+\frac{4}{(y+2)^2}.
$$
This relation together with (\ref{9.8}) leads to the following expressions
for the corresponding solutions of the $\mbox{KdV}_{3}$ hierarchy
\begin{equation}\label{9.21}
\begin{array}{l}
u_0=8q^6 \displaystyle\frac{y(y+4)(y^2+4y-4)}{(y+2)^6},\\\\
u_1=q^4 \displaystyle\frac{y(y+4)(y^2+4y-8)}{(y+2)^4}, \\ \\
u_2=-2q^2 \displaystyle\frac{y(y+4)}{(y+2)^2}.
\end{array}
\end{equation}
They represent coherent structures which propagate freely without
deformation. In the case of $y^{(3)}$ it determines a singular solution.

More general solutions of this type can be defined by increasing the
number of poles. Thus, one may take the
subspace $W\in\gr$ of boundary values of analytic functions $w=w(\lambda)$
on the disk $|\lambda|<1$, with the
possible exception of $n$ single poles at given real numbers
  ($0<|q_i|<1$ $i=1,\ldots,n$) and such that
$$
\lambda^2W\subset W \quad \mbox{and}\quad Res(w,q_i)=c_iw(-q_i),
$$
with $c_i>0$. The corresponding KdV wave function  reads
$$
\psi(\bt,\lambda)=\exp(\sum_{n\geq1}\lambda^{2n-1}t_{2n-1})
(1+\sum_{i=1}^n\frac{a_i(\bt)}{\lambda-q_i}),
$$
where the coefficients $a_i$ satisfy the system
$$
a_i(\bt)+c_i(\bt)\sum_{j=1}^n \frac{a_j(\bt)}{q_i+q_j}=c_i(\bt)
$$
with
$$
c_i(\bt):=c_i\exp(-2\sum_{n\geq1}q_i^{2n-1}t_{2n-1}) .
$$
By using again the functional $l(w)=w'(0)$,
the implicit equation $l(\psi)=0$
is now
$$
t_1\left( 1-\sum_{j=1}^n\frac{a_j(\bt)}{q_j}\right)
=\sum_{j=1}^n\frac{a_j(\bt)}{q_j^2}.
$$
It can be shown that the corresponding solutions of the
$\mbox{KdV}_{3}$ hierarchy represent composite structures
which decompose asymptotically into solutions of the form
(\ref{9.21}).

Another kind of solutions is obtained by considering
 the
subspace $W\in\gr$ of boundary values of analytic functions $w=w(\lambda)$
on the
 disk $|\lambda|<1$, with the
possible exception of a single pole at $\lambda=0$ and such that
$$
\lambda^2W\subset W \quad \mbox{and}\quad w(q)=cw(-q),
$$
for given $c>0$ and $q\in{\Bbb R}$ such that $-1<q<1$ . The wave function
is
$$
\psi(\bt,\lambda)=\exp(\sum_{n\geq1}\lambda^{2n-1}t_{2n-1})
(1+\frac{a(\bt)}{\lambda}),\;
a(\bt)=q\frac{c(\bt)-1}{c(\bt)+1},
$$
where
$$
c(\bt):=c\exp(-2\sum_{n\geq1}q^{2n-1}t_{2n-1}) .
$$
Solutions of the $\mbox{KdV}_{3}$ hierarchy can be derived by taking
 the functional
$$
l(w)=\frac{{\rm d} w}{{\rm d} \lambda}(q)+
c \frac{{\rm d} w}{{\rm d} \lambda}(-q),
$$
which verifies
$$
W\subset {\mbox Ker}(\lambda^2 l-q^2l).
$$
The implicit equation $l(\psi(\bt,\lambda))=0$
leads to
$$
\sum_{n\geq0} (2n+1)q^{2n}t_{2n+1}=\frac{1}{4q}\left(c(\bt)-
\frac{1}{c(\bt)}\right).
$$
By introducing the new variables
$$\begin{array}{l}
y:=2\sum_{n\geq 0} q^{2n+1} t_{2n+1}-\log c,\;\; x:=s_3,\\ \\
z:=4q^3 x+\log c+4\sum_{n\geq 2} nq^{2n+1}s_{2n+1},
\end{array}
$$
the equation reduces to
$$
y+z=-\sinh y.
$$
It defines  one implicit branch $y=y(z)$ which
satisfies
$$
\frac{{\rm d} y}{{\rm d} z}=-\frac{1}{1+\cosh y}.
$$
>From this relation and (\ref{9.8}) we get  the following expressions
for the associated solution of the $\mbox{KdV}_{3}$ hierarchy
$$
\begin{array}{l}
u_0=-\displaystyle\frac{4q^6}{\cosh ^6\frac{y}{2}}, \\\\
u_1=q^4\displaystyle\Bigl(1+ \frac{2}{\cosh ^2\frac{y}{2}}+
\frac{3}{\cosh ^4\frac{y}{2}}\Bigr) , \\ \\
u_2=-2q^2\displaystyle\Bigl(1+\frac{1}{\cosh ^2\frac{y}{2}}\Bigr).
\end{array}
$$
They again represent
coherent structures propagating  freely  and without
deformation.

The same strategy can be applied  for characterizing solutions of the
hidden Boussinesq
hierarchy. Let us first take the subspace $W\in\gr$ of boundary
values of functions $w=w(\lambda)$ analytic on the unit disk
$|\lambda|<1$, with the possible
exception of a single pole $q$ and such that
$$
\lambda^3W\subset W \quad \mbox{and}\quad
\frac{{\rm d} w}{{\rm d} \lambda}(0)=0.
$$
This subspace determines a wave function for the Boussinesq hierarchy
given by
$$
\psi(\bt,\lambda)=g(\bt,\lambda)\left( 1+\displaystyle\frac{q^2 t_1}
{(1+qt_1)(k-q)}\right).
$$
Consider now the functional
$$
l(w)=\frac{{\rm d}^2 w}{{\rm d} \lambda^2}(0),
$$
which obviously satisfies $\lambda^3\cdot l=0$. The equation
$l(\psi(\bt,\lambda))=0$
implies
$$
qt_1^2+2t_1-2qt_2=0,
$$
so that one finds the following explicit solution of the hidden Boussinesq
 hierarchy
$$\begin{array}{ll}
u_0=-\frac{12q^6}{(1+2q^2x)^3},& u_1=\frac{q^3}{(1+2q^2x)^{3/2}},\\
\noalign{\smallskip}
v_0=\frac{3q^4}{(1+2q^2x)^2},&v_1=\frac{3q}{(1+2q^2x)^{1/2}},
\end{array}
$$

Other solutions can be generated by starting with the same subspace $W\in
\gr$ and by taking the functional
$$
l(w)=\frac{{\rm d}^4 w}{{\rm d} \lambda^4}(0).
$$
In this case
$$
\lambda^3l(w)=24\frac{{\rm d} w}{{\rm d} \lambda}(0),
$$
so that $W\subset\mbox{Ker}(\lambda^3 l)$. The constraint
$l(\psi(\bt,\lambda))=0$
takes the form
$$
q^3t_1^4+4q^2t_1^3+4q(2+q^2t_2)t_1^2+8(1+q^2t_2)t_1-4q^3(2t_4+t_2^2)=0.
$$
A particular solution $t_1=b_1(\bs_2)$ of this equation is
$$
b_1=-\frac{1}{q}+\frac{1}{q}
\sqrt{-1-2q^2s_2+2\sqrt{1+2q^4s_4+2q^2s2+2q^4s_2^2}}.
$$
It can be seen that the corresponding solution,
as  a function of $x$ , is
globally defined on ${\Bbb R}$ only for $s_4>-\frac{1}{4q^4}$, otherwise
its domain is ${\Bbb R}-[-q^2(1+\sqrt{-1-4q^4 s_4}),-q^2(1-\sqrt{-1-4q^4
s_4})]$.

\end{document}